\documentclass[aps,showpacs,amsmath,dvips,10.50pt,showkeys,nofootinbib,preprintnumbers]{revtex4-2}

\usepackage{graphicx}   
\usepackage{color}      
\usepackage{braket}
\usepackage{tikz}
\usepackage{pgfplots}
\usepackage{orcidlink}
\usepackage{booktabs}
\def \be  {\begin{equation}}
\def \ee  {\end{equation}}
\def \ee  {\end{equation}}
\def \bea {\begin{eqnarray}}
\def \eea {\end{eqnarray}}

\newcommand{\nn}{\nonumber}

\usepackage{titlesec}
\titleformat{\section}[block]{\normalfont\bfseries}{\thesection.}{1em}{}
\titleformat{\subsection}[block]{\normalfont\bfseries}{\thesubsection}{1em}{}

\begin{document}
\title{Charmed Meson Structure across Crossover from SU($4$) Polyakov Quark Meson Model with Isospin Asymmetry}
\author{Abdel Magied DIAB \orcidlink{0000-0002-5603-1952}}
\email{a.diab@eng.mti.edu.eg}
 \affiliation{Physics and Engineering Mathematics Department, Faculty of Engineering, Modern University for Technology and Information (MTI), 11571 Cairo, Egypt.}
\date{\today}
\begin{abstract}%
The Polyakov Quark Meson (PQM) model is extended to SU($4$) flavor symmetry by incorporating the charm quark and introducing a finite isospin asymmetry. This model incorporates the light, strange, and charm chiral condensates, along with the Polyakov-loop variables, to describe the confinement--deconfinement phase transition in a thermal and dense QCD medium. The inclusion of the charm quark condensate enhances the capability of the SU($4$) PQM model to explore the spatial and thermal resolution of the chiral phase structure, particularly in the crossover and high-temperature regimes. We construct the QCD phase diagram ($T/T_\chi-\mu_I/m_\pi$) plane, indicating a decrease in the pseudo-critical temperature as the isospin chemical potential increases and explore thermodynamic quantities related to the QCD equation of state at very high temperatures.  Fluctuations of quark flavors, conserved charges and baryon-charm correlations are  studied across a wide temperature range. The SU($4$) PQM model exhibits good qualitative agreement with lattice QCD calculations. Additionally, we calculate the meson mass spectrum at zero and finite temperature, showing that the charm sector remains thermally stable over a wide temperature range. Overall, this study highlights the capability of the SU($4$) PQM model to describe key features of QCD matter at high temperatures and its relevance to heavy-ion collisions and astrophysical studies.  
\end{abstract} 

\maketitle
\section{Introduction \label{intro}}
Recent progress in ultra-relativistic heavy-ion experiments--such as the Relativistic Heavy Ion Collider (RHIC) at Brookhaven National Laboratory (BNL), the Large Hadron Collider (LHC) at CERN, the upcoming Compressed Baryonic Matter (CBM) experiment at the Facility for Antiproton and Ion Research (FAIR), and the Nuclotron-based Ion Collider fAcility (NICA) at the Joint Institute for Nuclear Research (JINR) --has significantly advanced the investigation of strongly interacting matter. Quantum Chromodynamics (QCD), the fundamental theory of strong interactions among quarks and gluons, predicts that at high temperatures and baryon densities, hadronic matter undergoes a phase transition into a deconfined and chirally symmetric phase known as the quark-gluon plasma (QGP)  \cite{Rischke:2003mt,Fukushima:2010bq,Muller:1994rb,Svetitsky:1985ye,Rapp:2009my,McLerran:1981pb}.

In recent decades, effective low-energy QCD models have been developed investigate different features of the QCD phase transition. These approaches incorporate key concepts of spontaneous chiral symmetry breaking and color-charge deconfinement \cite{Gasiorowicz:1969kn,Meissner:1987ge,Vafa:1983tf}. The chiral symmetry is a fundamental symmetry in QCD in the limit of vanishing quark masses \cite{Gasiorowicz:1969kn,Meissner:1987ge}. For finite quark masses, chiral symmetry becomes spontaneously broken, with the chiral condensate serving as the associated order parameter. When quark masses are small (up $u$- and down $d$-quarks), chiral symmetry becomes explicitly broken, resulting in the pseudo-Goldstone bosons acquiring finite masses, for example, the light pseudoscalar mesons  as pions and kaons. For heavier quarks, such as strange and charm quarks, the explicit breaking is stronger, leading to the formation of hidden and open charmed mesons such as $D$-resonances and $\chi_{c0}$. At high temperature or density, the chiral condensate melts, signaling the restoration of chiral symmetry and the onset of meson state degeneracy.

Low-energy QCD approaches have been widely employed to study the QCD phase structure. These approaches, such as chiral perturbation theory~\cite{Espriu:2020dge}, Dyson-Schwinger Equations (DSEs) \cite{Zhao:2019ruc,Fischer:2018sdj,Ayala:2011vs}, hadron resonance gas and other statistical models, for instance, \cite{Karsch:2003vd,Ejiri:2005wq,Braun-Munzinger:2003htr}, the Polyakov Nambu-Jona-Lasinio (PNJL) model~\cite{Zhao:2020xob,Buballa:2003qv,Zhou:2025uxv,Chu:2017bmr}, the Polyakov linear sigma model (PLSM) or Polyakov quark meson (PQM) model~\cite{Schaefer:2008hk,Tawfik:2014gga,Tawfik:2021eeb,Fukushima:2003fw} have been widely applied. In the world of $N_f$ quark flavors with $u$- and $d$-quarks, i.e., SU($2$) flavor symmetry \cite{Gallas:2009qp,Parganlija:2010fz,Janowski:2011gt}, pions are formed, and the light quark condensate is investigated. The extension to SU($3$) symmetry includes the strange $s$-quark, allowing for the investigation of the strange condensate, nonet meson masses, and the QCD phase transition~\cite{tHooft:1976rip,tHooft:1986ooh,Parganlija:2012fy}. Further extension to SU(4) flavor symmetry incorporates the charm quark and its associated condensate, making it possible to analyze the charmed meson states in a thermal and dense QCD medium \cite{Eshraim:2014eka,Eshraim:2013exr,Diab:2016iig,AbdelAalDiab:2018hrx}. The validity of the PQM model has been confirmed by extensive comparisons with lattice QCD simulations, demonstrating key features of QCD thermodynamics and phase transitions~\cite{Tawfik:2014uka,Tawfik:2016edq,Tawfik:2019kaz}, including the non-extensive effects on the QCD EoS \cite{Diab:2024vwh}, SU($3$) isospin asymmetry~\cite{Tawfik:2019tkp}, SU($4$) symmetry and QCD thermodynamics~\cite{Diab:2016iig,AbdelAalDiab:2018hrx}, and viscous properties \cite{Tawfik:2016edq}. Further extensions studied the chiral symmetry containing the vector meson states at finite density \cite{Tawfik:2014gga} and magnetic field~\cite{Tawfik:2019rdd}. Other studies explored the QCD EoS in the presence of a magnetic field~\cite{Tawfik:2016ihn,Tawfik:2016lih,Ezzelarab:2015tya} and magnetic catalysis \cite{Tawfik:2016gye,Tawfik:2017cdx}.

One of the main targets of the research programs at  NICA (JINR) \cite{Toneev:2007yu} is to investigate  in-medium properties of the QCD phase diagram at high density and moderate temperatures by colliding heavy ions, exploring the onset of deconfinement, locating the chiral symmetry restoration boundary, and searching for the critical endpoint (CEP). The created matter  will naturally exhibit finite isospin asymmetry, introducing a nonzero isospin chemical potential ($\mu_I$) into the system. Theoretical approaches such as the PQM model and lattice QCD provide essential guidance for analyzing experimental data, accurately mapping the  chiral QCD phase structure, and identifying the CEP in iso-asymmetric QCD matter. The motivation behind the current manuscript is to utilize and examine the inclusion of finite isospin asymmetry and the charm chiral condensate on the chiral QCD phase structure, the chiral phase diagram of QCD matter, thermodynamic observables, diagonal and off-diagonal fluctuations of baryon-charm interactions, and the mass spectrum of (pseudo)scalar meson states at finite temperature $T$, particularly at very high temperatures on the order of $1~\text{GeV}$. The inclusion of the charm chiral condensate significantly improves the capability of the PQM model to evaluate the QCD phase structure, especially at high temperatures. This enhancement is relevant not only for particles physics but also for modeling the equation of state (EoS) in hyperonic matter, Bose-Einstein condensation (BEC), color superconductivity (CSC), and the EoS for the  properties of neutron stars (NS) and merger remnants~\cite{Steiner:2004fi,Akmal:1997ft}.

The manuscript is organized as follows. Section~\ref{Model1} outlines the structure of the SU($4$) PQM model in finite isospin asymmetry. Section~\ref{resultsPQM} focuses on  the analysis of the SU($4$) PQM model in a thermal and dense QCD medium. The PQM order parameters and the chiral QCD phase diagram are discussed in subsection~\ref{Orderparameter}, while thermodynamic observables, the EoS of QCD matter, fluctuations and correlations of conserved charges over a wide temperature range are explored in subsection~\ref{QCDthermo}. subsection~\ref{masscurve} examines the meson mass spectrum at both zero and finite temperatures. Last but not the least, section~\ref{conclusion} provides a summary and conclusion.

\section{Model and Formalism   \label{Model1}}
The SU($N_f$) PQM model addresses the phase transition between the confinement and deconfinement phases,  the  chiral Lagrangian is incorporating with the Polyakov-loop potential as $\mathcal{L}_{\mathrm{PQM}} = \mathcal{L}_{\mathrm{chiral}} - \mathcal{U}(\phi, \bar{\phi}, T)$. The chiral Lagrangian $\mathcal{L}_{\mathrm{chiral}}$ for ($N_f=4$) quark flavors and  ($N_c = 3$) color degree of freedom is given as follows:
\bea
\mathcal{L}_{\mathrm{chiral}}&=&\mathcal{L}_{\bar{\psi}\psi} + \mathcal{L}_{\mathcal{M}} + \mathcal{L}_U(1)_A +   \mathcal{L}_{\mbox{emass}}, \label{ChrLAge}
\eea
where,
\bea
\mathcal{L}_{\bar{\psi}\psi} &=&  \sum_f \overline{\psi}_f \Big[i\gamma^{\nu} D_{\nu}-g\,T_a(\sigma_a+i \gamma_5 \pi_a)\Big] \psi_f,\nn \\ 
\mathcal{L}_{\mathcal{M}}  &=& \mathrm{Tr}\Big[\partial_{\mu}\mathcal{M}^{\dag}\partial^{\mu}\mathcal{M}-m^2 \mathcal{M}^{\dag} \mathcal{M}\Big]  - \lambda_1 \Big(\mathrm{Tr}\big[\mathcal{M}^{\dag} \mathcal{M}\big]\Big)^2 - \lambda_2 \mathrm{Tr}\Big[\mathcal{M}^{\dag} \mathcal{M}\Big]^2 + \mathrm{Tr}[H(\mathcal{M}+\mathcal{M}^{\dag})],\nn \\ 
\mathcal{L}_U(1)_A &=& \mathcal{C} \Big(\mathrm{Det}\Big[\mathcal{M}\Big] \mathrm{Det}\big[\mathcal{M}^{\dag}\Big]\Big),\nn \\ 
\mathcal{L}_{\mbox{emass}} &=& - 2\mathrm{Tr}\Big[\varepsilon \Big(\mathcal{M}^{\dag} \mathcal{M}\Big)\Big]. 
\eea
Here, the first term on the right-hand side (rhs) of Eq.~(\ref{ChrLAge}) represents the Lagrangian density for the baryonic (fermionic) field $\psi_f$, where $D_{\nu}$, $\nu$, $\gamma^{\nu}$, and $g$ denote the covariant derivative, the Lorentz index, the chiral spinors, and the Yukawa coupling constant, respectively. The index $f$ refers to the quark flavors, given as $f = [(l = u, d), s, c]$. The remaining terms on the rhs of Eq.~(\ref{ChrLAge}) correspond to the contributions from the mesonic (bosonic) field $\mathcal{M}$. The generator operator $\mathcal{M}$ is represented by a complex matrix~\cite{Lenaghan:2000ey,Roder:2003uz} as,
\bea 
\bar{\mathcal{M}} =  \sum_{a=0}^{N_f^2 - 1} T_a ( \bar{\sigma_a}+i  \bar{\pi_a})
=\frac{1}{\sqrt{2}}
   \begin{pmatrix}
\frac{(\sigma_\ell +a_0^0)+i (\eta_\ell +\pi^0)  }{\sqrt{2}}& a_0^+ +i  \pi^+& \kappa^+ +i K^+& D_0^0 +i\, D^0 \\
a_0^- +i \pi^- &\frac{(\sigma_\ell -a_0^0)+i (\eta_\ell -\pi^0)  }{\sqrt{2}}& \kappa^0 +i  K^0 & D_0^-+i\, D^- \\
\kappa^-  +i  K^-& \overline{\kappa}^0 +i  \overline{K}^0 &\sigma_S +i\eta_s& D_{S,0}^- +i  D_{S}^- \\
\overline{D}_0^0 +i  \overline{D}^0   & D_0^+ +i  D^+ & D_{S,0}^+ +i   D_{S}^+   &\chi_{c0}+i\eta_C
   \end{pmatrix}. \label{MesonMAtrix}
\eea

 \begin{table}[htb]
 \begin{center}
\begin{tabular}{l|p{13.cm}}  
 \hline  \hline
 Pseudoscalar   &
--The pion triplet states $\pi^\pm=(\pi_1\pm i \pi_2)/\sqrt{2},\pi^0=\pi_3$, \\
sectors&--four kaon states  $K^\pm=(\pi_4\pm i \pi_5)/\sqrt{2},   K^0 =(\pi_6+ i \pi_7)/\sqrt{2}\quad \quad$ and $\overline{K}^0=(\pi_6 - i \pi_7)/\sqrt{2}$, \\
&--isoscalar fields $\eta_{\ell,s}$  mix and generate the physical fields  $\eta$ and $\eta^\prime$ \cite{Parganlija:2012fy, Parganlija:2012ogz, Schaefer:2008hk}. \\
$J^{PC} = 0^{-+}$ &--open charmed $D$ resonance  $D^\pm =(\pi_{11} \pm i \pi_{12})/\sqrt{2},  D^0 =(\pi_{9} - i \pi_{10})/\sqrt{2}, \overline{D}^0=(\pi_{9} + i \pi_{10})/\sqrt{2}$, \\
&--open strange-charmed $D_S^\pm=(\pi_{13} \pm i \pi_{14})/\sqrt{2}$, and \\
&--hidden charmed $\eta_C=(\pi_0-\sqrt{3} \pi_{15})/2$.\\
  \hline  \hline
 Scalar   &
--The isotriplet of $\vec{a}_0$ meson  $a_0^\pm=(\sigma_1 \pm i \sigma_2)/\sqrt{2}, a_0^0=\sigma_3$ \\ 
sectors&--four kaon states $\kappa ^\pm=(\sigma_4 \pm i \sigma_5)/\sqrt{2}$, $\kappa^0 =(\sigma_6+ i \sigma_7)/\sqrt{2}$ and $\overline{\kappa}^0=(\sigma_6 - i \sigma_7)/\sqrt{2}$, \\
&--isoscalar fields $\sigma_{\ell,s}$ mix and generate the physical fields $f_0$ and $\sigma$ \cite{Parganlija:2012fy, Parganlija:2012ogz, Schaefer:2008hk}, \\
$J^{PC} = 0^{++}$&--open charmed $D_0$ resonance  $D_0^\pm =(\sigma_{11} \pm i \sigma_{12})/\sqrt{2}, D_0^0 =(\sigma_{9} - i \sigma_{10})/\sqrt{2},\overline{D}_0^0=(\sigma_{9} + i \sigma_{10})/\sqrt{2}$, \\
&--open strange-charmed $D_{S,0}^\pm=(\sigma_{13} \pm i \sigma_{14})/\sqrt{2}$, and \\
&--hidden charmed $\chi_{c0}=(\sigma_3-\sqrt{3} \sigma_{15})/2$.
\\
  \hline  \hline
\end{tabular}
\caption{A summary of the field contents of meson sectors. \label{TableSummrya}}
\end{center}
 \end{table}
Within the U($4$) algebra framework with the pseudoscalar ($\pi_a, J^{\mathcal{PC}=0^{-+}}$) and scalar ($\sigma_a, J^{\mathcal{PC}=0^{++}}$) sectors, the generators $T_a= \hat{\lambda}_a/2$ can be derived from Gell-Mann matrices $\hat{\lambda}_a$  \cite{Weinberg:1972kfs} , where the index $a$ runs as $a=0,\cdots,\,15$ (see App.\ref{AppA}). The fields  $\sigma_a,\, \pi_a$ denote the scalar ($J^{PC} = 0^{++}$) and pseudoscalar  ($J^{PC} = 0^{-+}$) nonet meson fields, respectively \cite{Lenaghan:2000ey, Roder:2003uz}. For sake of clarity, Eq. (\ref{MesonMAtrix}) illustrates the meson sectors as given in Tab. \ref{TableSummrya}. In vacuum phenomenology, the algebraic group $U(4)$ symmetry is decomposed as $U(4) \approx SU(4) \times U(1)_A$, where $U(1)_A$ is represented by the anomaly term $\mathcal{L}_{U(1)_A} = \mathcal{C} \left(\mathrm{Det}[\mathcal{M}] + \mathrm{Det}[\mathcal{M}^{\dagger}]\right)$. Moreover, one can assume that $H \neq 0$, $\mathcal{C} \neq 0$, and $\lambda \neq 0$. This implies that the chiral Lagrangian leads to nonzero quark masses, and the anomaly term is explicitly broken. As a consequence of chiral symmetry is spontaneous broken, the mean value of the mesonic field, $\langle \mathcal{M} \rangle = T_0 \sigma_0 + T_3 \sigma_3 + T_8 \sigma_8 + T_{15} \sigma_{15}$, must carry specific quantum numbers, with only $h_0$, $h_3$, $h_8$, and $h_{15}$ being non-vanishing. To address finite isospin asymmetry, where the quark flavor masses are not completely degenerate (i.e., $m_u \neq m_d,\;m_s,\;m_c$), we utilize an orthogonal basis transformation. This transformation converts the condensates from the original basis, $\sigma_0$, $\sigma_3$, $\sigma_8$, and $\sigma_{15}$, to the pure up ($\sigma_u$), down ($\sigma_d$), strange ($\sigma_s$), and charm ($\sigma_c$) flavor basis, respectively. This conversion can be expressed as,
\bea
   \begin{pmatrix}
\sigma_u \\ \sigma_d \\ \sigma_s \\ \sigma_c
   \end{pmatrix}
=
  \begin{pmatrix}
\frac{1}{\sqrt{2}} & 1 & \frac{1}{\sqrt{3}}  & \frac{1}{\sqrt{6}} 	\\
\frac{1}{\sqrt{2}} & -1 & \frac{1}{\sqrt{3}}  & \frac{1}{\sqrt{6}}\\
\frac{1}{2}  & 0 & \sqrt{\frac{2}{3}} & \frac{1}{2\sqrt{3}}	\\
\frac{1}{2}  & 0 & 0 & -\frac{\sqrt{3}}{2}
  \end{pmatrix}
   \begin{pmatrix}
\sigma_0 \\ \sigma_3 \\ \sigma_8 \\ \sigma_{15}
\end{pmatrix}
\eea
The chiral symmetry is explicitly broken by ($4\times 4$) matrix, $H= T_a h_a$, the explicitly symmetry breaking terms, $h_a$ can be determined from the partially conserved axial current (PCAC) relations \cite{Lenaghan:2000ey,Tawfik:2019tkp}. The parameters of explicit symmetry breaking for light ($l=u-,d-$quark), strange ($s-$quark) and charm ($c-$quark) are given as,
\bea
h_l = f_\pi m_\pi^2, \quad \quad h_s = \sqrt{2} f_K m_K^2 -\frac{f_\pi m_\pi^2}{\sqrt{2}}, \quad \quad h_c = \sqrt{2} f_D m_D^2 - \frac{f_\pi m_\pi^2}{\sqrt{2}}.
\eea 
With this regard, we utilize the PQM model to analyze the QCD phase structure in  finite isospin asymmetry, we proceed as follows: Firstly, we investigate the impacts of isospin asymmetry to distinguishable between the light condensates of $u$- and $d$-quarks. This arises due to the breaking of  isospin asymmetry, where $\sigma_3 = a_0^0= \big(f_{K^\pm}-f_{K^0}\big)\neq 0$ \cite{Tawfik:2019tkp}. Secondly, we define the explicit symmetry breaking for light quarks $(l=u,d)$ as $h_u = h_l+h_3$ and $h_d = h_l-h_3$, where the term $h_3$ representing explicit symmetry breaking is given by $h_3 = m_{a0}^2 \big( f_{K^\pm}-f_{K^0}\big)$ \cite{Tawfik:2019tkp}. Finally, the term $-2\mathrm{Tr [\varepsilon (\mathcal{M}^{\dag} \mathcal{M})]}$ in Eq. (\ref{ChrLAge}), where $\varepsilon$ represents the direct contribution of bare quark to nonet meson masses, 
\bea \varepsilon =
   \begin{pmatrix}
   \varepsilon_u &0 & 0& 0\\
   0&\varepsilon_d & 0 & 0 \\
   0 &0& \varepsilon_s &0 \\
   0 &0 &0 & \varepsilon_c
      \end{pmatrix},
\eea
For sake simplicity, one can set $\varepsilon_u = \varepsilon_d = \varepsilon_s = 0$~\cite{Parganlija:2012fy,Eshraim:2014eka}. The used parameters $m^2$, $h_u$, $h_d$, $h_s$, $h_c$, $\lambda_1$, $\lambda_2$, and $\mathcal{C}$ are expressed in terms of $m_\sigma$ \cite{Schaefer:2008hk}. In the current manuscript, the relevant  parameters assume for $m_\sigma=800~$MeV. \cite{Schaefer:2008hk}. The Polyakov-loop potential, $\mathcal{U} (\phi, \overline{\phi}, T)$, incorporates  degrees of freedom of color-charges and the dynamics of the color gluon-quark interaction to the chiral Lagrangian of the PQM model. The potential is estimated by calculating the thermal expectation value of a color--traced Wilson loop in temporal space \cite{Susskind:1979up, Polyakov:1978vu} as $\phi=\mbox{Tr}_c[\mathcal{L}]/N_c$ and its hermitian conjugate $\overline{\phi}=\mbox{Tr}_c[L^\dagger]/N_c$, where $\mathcal{L}$ and $\mathcal{L}^\dagger$ are matrices in the representation of SU($N_c$) color-charge gauge group as \cite{Shifman:2001ck, Fukushima:2003fw,Ratti:2005jh}, 
\bea
\mathcal{L}(\vec{x}) = \mathcal{P} \exp{\Big[i\int_0^{1/T} d\tau  A_0(\tau, \vert{x})\Big]},
\eea
where $A_0$ and $\mathcal{P}$ are the temporal component and the path ordering of the gluon gauge field, respectively. Various functional forms of the Polyakov-loop potential exist \cite{Roessner:2006xn,Scavenius:2002ru,Pisarski:2000eq, Schaefer:2008hk,Lo:2013hla}; in this work, we adopt the Polyakov-loop potential proposed by Fukushima \cite{Fukushima:2008wg}, inspired by a strong-coupling interaction
\bea
\mathcal{U}_{\mbox{Fuku}} (\phi, \bar{\phi}, T)  = -b\;T\Big[54 e^{-a/T} + \ln{\Big(1- 6\bar{\phi} \phi - 3(\bar{\phi} \phi )^2 + 4 (\bar{\phi}^3+ \phi^3 )\Big)} \Big], \label{FukushimaPot}
\eea
where $a=664.0~$MeV and $b=196.2~$MeV are fixed parameters. The parameter $a$  controls the confinement transition in pure gauge, while b related to the mixing of the chiral and deconfinement transition \cite{Fukushima:2008wg}. The expectation value of the Polyakov loops is finite and approaches unity (indicating the deconfined phase) at high temperatures, while it diminishes to zero or takes small values at low temperatures (indicating the confined phase) \cite{Schaefer:2009ui}. Hence, this order parameter serves as a reliable indicator of the confinement-deconfinement phase transition. 

The grand canonical partition function $\mathcal{Z}$ which is given in dependence of the temperature $T$ and the chemical potentials of $f$-th quark flavor $\mu_f$ is defined by the path integral over all fermions, anti-fermions, and bosons. In mean-field approximation \cite{kapusta2007finite},  all fields are treated as averages in space ($\vec{x}$) and imaginary time ($\tau$) as,
\bea
\mathcal{Z} (T,\mu_f) &=&  \mbox{Tr} \exp \Big[ \sum_f \frac{  \mu_f \hat{\mathcal{N}}_f -\hat{\mathcal{H}}}{T} \Big]  \nn \\ &=& \int  \prod_a \mathcal{D}_{\sigma_a} \mathcal{D}_{\pi_a} \int \mathcal{D}_{\psi_f} \mathcal{D}_{\overline{\psi}_f}  \exp{\Big[ d^4\vec{x} \Big( \mathcal{L} + \sum_f \mu_f \overline{\psi}_f  \gamma^0  \psi_f \Big) \Big]},
\eea
where $\mathcal{N}_f \; \mbox{and} \; \hat{\mathcal{H}}$ are defined by the quark number density and the chiral Hamiltonian density, respectively. The chemical potentials $\mu_f$ are related to the conserved quantum numbers of baryon ($B$), electric charge ($Q$), isospin ($I$), strangeness ($S$) and charm ($C$) number for each of the quark flavors,  
\bea
\mu_u &=& \frac{1}{3}\mu_B + \frac{2}{3}\mu_Q +\frac{1}{2} \mu_I, \quad\quad 
\mu_d = \frac{1}{3}\mu_B - \frac{1}{3}\mu_Q -\frac{1}{2} \mu_I, \nn \\ 
\mu_s &=& \frac{1}{3}\mu_B - \frac{1}{3}\mu_Q -  \mu_S,\quad\quad 
\mu_c = \frac{1}{3}\mu_B + \frac{2}{3}\mu_Q +\mu_C . \label{chemicalPota}
\eea
The exchanges between particles and antiparticles are incorporated through the grand-canonical partition function ($\mathcal{Z}$), which is constructed from the free energy density as,
\bea
\mathcal{F}_{PQM}(T, \mu_f) =-T\, \frac{\ln{\mathcal{Z}}(T,\mu_f)}{V}= \mathcal{U}(\sigma_u,\, \sigma_d,\, \sigma_s,\, \sigma_c)+ \mathcal{U}_{\bar{\psi}\psi}(T, \mu _f) +\,\mathcal{U}_{\mbox{Fuku}} (\phi, \bar{\phi}, T). \label{LSMPOT}
\eea 
The first term on rhs of Eq. (\ref{LSMPOT}) represents the tree-level mesonic potential. It accounts for the contribution of the pure multiplet of the nonet meson states with a basis of light ($\sigma_u$, $\sigma_d$), strange ($\sigma_s$), and charm ($\sigma_c$) condensates. The purely bosonic (mesonic) potential is given as,
\bea
\mathcal{U}(\sigma_u,\, \sigma_d,\, \sigma_s,\, \sigma_c) &=& -\frac{h_u}{2} \sigma_u-\frac{h_d}{2} \sigma_d - h_s \sigma_s - h_c \sigma_c  + \frac{m^2}{2}\;\Big(\frac{\sigma^2_u}{2}+\frac{\sigma^2_d}{2}+\sigma^2_s+\sigma^2_c\Big)-\frac{\mathcal{C}}{4}\, \sigma_u \sigma_d \sigma_s \sigma_c  \nn \\ &+& \varepsilon_c \sigma^2_c + \frac{\lambda_1}{16}\Big(\sigma_u^2+\sigma_d^2+ 2\sigma_s^2+2\sigma_c^2\Big)^2 + \frac{\lambda_2}{16}\Big(\sigma_u^4+\sigma_d^4+ 4\sigma_s^4+4\sigma_c^4\Big)^2.
\eea
The second term in rhs of  Eq. (\ref{LSMPOT}) represents the fermion contributions to the PQM model \cite{Fukushima:2008wg,Fukushima:2003fw} as,
\bea
\mathcal{U}_{\bar{\psi}\psi}(T, \mu _f)&=& \mathcal{U}_{\bar{\psi}\psi}^{\mbox{vac}} + \mathcal{U}_{\bar{\psi}\psi}^{\mbox{the}} \nn \\ &=& -2N_c\,\sum_f \int \frac{d^3k}{(2\pi)^3}\Big\{\; E_f + \frac{T}{N_c}\;\Big(\ln{[g_D^{(+)}(T, \; \mu_f)]} + \ln{[g_D^{(-)}(T, \; \mu_f) ]} \Big)\Big\}, \label{qaurkPot}
\eea  
the first term in the rhs of Eq. (\ref{qaurkPot}) corresponds to the fermionic vacuum contribution, this term is usually neglected in calculations, Moreover, one can regularise the dimensions to avoid the diverging integral as \cite{Skokov:2010sf,Gupta:2011ez},
\bea
\mathcal{U}_{\bar{\psi}\psi}^{\mbox{vac}} \equiv  \mathcal{U}_{\bar{\psi}\psi}^{\mbox{reg}} (\Lambda) = -\frac{N_c}{8\pi^2} \sum_f m_f^4 \log{\Big[\frac{m_f}{\Lambda}\Big]},
\eea
where $\Lambda$ is defined as a regularisation scale parameter. While, the second term in the rhs of Eq. (\ref{qaurkPot}) gives the fermionic contribution in thermal and dense QCD-medium, whereas $g_D^{(+)}(T, ; \mu_f)$ and $g_D^{(-)}(T, ; \mu_f)$ are defined as the Fermi-Dirac distribution functions in the presence of expectation values of Polyakov loop variables, following widely used notations as,
\bea
g_D^{(+)}(T, \; \mu_f) &=&   1+ 3\Big(\phi\,+\overline{\phi}\;e^{-\frac{E_f^{(+)}}{T}}\Big) e^{-\frac{E_f^{(+)}}{T}}+e^{-3 \frac{E_f^{(+)}}{T}}, \label{Modfermi}
\eea
where  $E_f^{(\pm)}=E_f(k) \mp \mu_f$ represent the energy-momentum dispersion relations, with the upper sign applying to quarks and the lower sign to antiquarks. The energy-momentum dispersion relations for quark flavors are defined as, $E_f(k)=(k^2+m_f^2)^{1/2}$, where $m_f$ denotes the constituent quark masses as,
\bea
m_u= g\frac{\sigma_u}{2}, \quad \quad
m_d= g\frac{\sigma_d}{2}, \quad \quad
m_s = g\frac{\sigma_s}{\sqrt{2}}, \quad \quad 
m_c = g\frac{\sigma_c}{\sqrt{2}}.
\eea
The term $g_D^{(-)}(T, ; \mu_f)$ can be evaluated by substituting $E_f^{(+)} $ with $E_f^{(-)}$ and replacing the Polyakov-loop variable $\phi$ with its conjugate $\overline{\phi}$, and vice versa. As a result of the chiral symmetry is spontaneously broken, the PQM Lagrangian undergoes a shift in the meson fields $\sigma_f$ by their vacuum expectation values, denoted as $\sigma_f \longrightarrow \sigma_f + \langle \sigma_{f0} \rangle$. The values of quark chiral condensates in vacuum are calculated in terms of decay widths  of pion, kaon, and $D$-meson by using the partially conserved axial-vector current (PCAC) relations,  $\overline{\sigma}_{\ell o}  =  f_\pi = 92.4~$MeV, $\overline{\sigma}_{so} =(2   f_K -   f_\pi)/\sqrt{2}= 94.5~$MeV, and $\overline{\sigma}_{co} =(2 f_D - f_\pi)/\sqrt{2} = 293.87~$MeV. We use the experimental values \cite{ParticleDataGroup:2018ovx, Aoki:2016frl}, $f_{\pi^{\pm}} = f_{\pi^0} =92.4\; \mathrm{MeV}$, $f_{K^{\pm}} =113\; \mathrm{MeV} , \;f_{K^0} =113.453 \;\mathrm{MeV}$ and $f_D= 254~$MeV are the decay constants of pion, kaon and $D$ meson states. Within the mean-field approximation (MFA),  one can examine the thermodynamic observables that characterize the chiral QCD phase structure of quark-hadron matter as functions of independent parameters such as temperature, chemical potentials, magnetic field and so on. These observables are derived by examining the thermal behavior of the order parameters of the PQM model. Within this framework, the quark chiral condensates $\overline{\sigma}_u$, $\overline{\sigma}_d$, $\overline{\sigma}_s$, and $\overline{\sigma}_c$, as well as the expectation values of the Polyakov-loop variables $\overline{\phi}$ and $\overline{\bar{\phi}}$ are determined by locating the global minimum of the real part of the PQM thermodynamic  free energy density  $\mathcal{R} \;[\mathcal{F}_{PQM}(T, \mu_f)]$, as given by Eq. (\ref{LSMPOT}) to the relevant order parameters at the saddle point,
\bea
\frac{\partial\mathcal{F}_{PQM}}{\partial\sigma_u}=\frac{\partial\mathcal{F}_{PQM}}{\partial\sigma_d}= \frac{\partial\mathcal{F}_{PQM}}{\partial\sigma_s} =\frac{\partial\mathcal{F}_{PQM}}{\partial\sigma_c}= \frac{\partial\mathcal{F}_{PQM}}{\partial\phi} = \frac{\partial\mathcal{F}_{PQM}}{\partial\bar{\phi}}\Bigg\vert_{\mbox{min}} = 0. \label{saddleEq}
\eea 
The Polyakov loop is a complex-valued matrix, which renders the PQM free energy density in Eq. (\ref{LSMPOT}) a functional of complex variables. Notably, the physical observables derived from the imaginary part of this function vanish, ass] discussed in Ref~\cite{Sasaki:2006ww}.   This involves solving a set of equations simultaneously.  The thermal evolution of the order parameters is analyzed by using the gap equations and by tracking the approach towards the saddle point.
\section{Results and Discussion \label{resultsPQM}} 
 This section investigates the properties of the quark-hadron phase structure using the SU($4$) PQM model. The analysis highlights several thermodynamic observables, such as chiral condensates, deconfinement order parameters, pseudo-critical temperatures, and other thermodynamic quantities, compared with lattice QCD simulations. Additionally, we examine the quark susceptibilities, fluctuations and correlations of conserved charges associated with the charm quantum number. Last but not least, the mass spectrum of nonet meson states is introduced within the framework of the SU($4$) PQM model.

\begin{figure}[htb]
\centering{
\includegraphics[width= 16.5 cm, height=5.5 cm, angle=0  ]{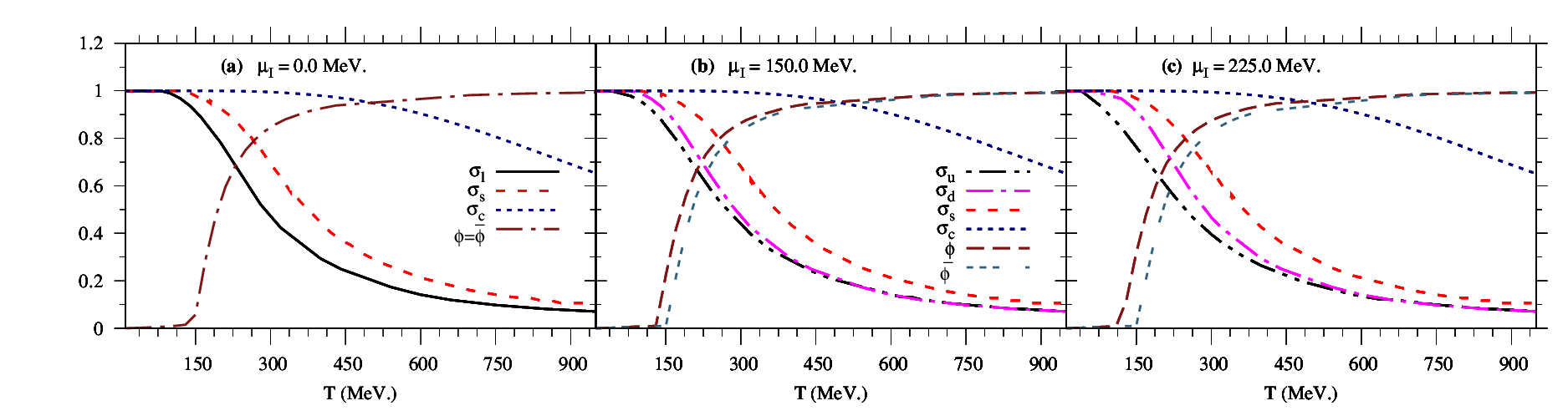}
\caption{ (Color online) Left panel (a) illustrate the thermal behaviors of the PQM order parameters ($\sigma_f/\sigma_{f_0}$, $\phi$, and $\bar{\phi}$) at  $\mu_I=0.0$. The middle and right panels (b) and (c) show  the same as (a) but at $\mu_I=150.0~$MeV and $\mu_I=225.0~$MeV, respectively.}
\label{Fig-orderParameters-T}} 
\end{figure} 
\subsection{PQM Order Parameters and Pseudo-Critical Temperatures \label{Orderparameter}}
 It is helpful to present several calculations to illustrate the impact of isospin asymmetry on the chiral phase transition and deconfinement order parameters in the PQM model for $N_f = 2+1+1$ quark flavors. To achieve a reliable differentiation between $u$- and $d$-quark condensates, it is necessary to estimate the influence of finite isospin asymmetry on the SU($4$) PQM order parameters. To this end, we begin with the real part of the PQM thermodynamic free energy density in the mean-field approximation (MFA), Eq.~(\ref{LSMPOT}). This term should be globally minimized at a saddle point with respect to the corresponding mean field, as outlined in the gap equations, Eq.~(\ref{saddleEq}). The solutions of the gap equations provide a complete set of equations  that allow for illustrating the evolution of the chiral condensates $\sigma_f$ with $[f=(l=u,d),s,c]$ quark flavors, as well as the order parameters of Polyakov-loop potential $\phi$ and $\bar{\phi}$, in the thermal and dense QCD medium.
 
Figure~\ref{Fig-orderParameters-T} depicts the normalized quark chiral condensates $\sigma_f/\sigma_{f_0}$ to their values in vacuum and the order parameters of the Polyakov-loop variables($\phi$ and $\bar{\phi}$) as a function of temperature at different values of isospin ($\mu_I$) and baryon ($\mu_B$) chemical potentials: the left panel (a) at $\mu_I = 0.0$, the middle panel (b) at $\mu_I = 150.0~\text{MeV}$, and the right panel (c) at $\mu_I = 225.0~\text{MeV}$, respectively. In the current calculations, we assumed that $\mu_B = 2.0\mu_I$, and other chemical potentials for conserved charges are set to zero. The left panel (a) displays the normalized expectation values of the nonstrange ($\sigma_{\ell}$, solid curve), strange ($\sigma_s$, dashed curve), and charm ($\sigma_c$, dotted curve) chiral condensates, as well as the order parameter for the Polyakov-loop variables $\phi = \bar{\phi}$ (dotted-dashed curve), at zero isospin chemical potential ($\mu_I = 0.0$). The nonstrange condensate is defined as $\sigma_\ell = (\sigma_u + \sigma_d)/2$. In the hadronic phase, when the temperature is below the pseudo-critical temperature $T_\chi$, the values of $\sigma_{f}/\sigma_{f_0}$ start at unity and remain unchanged until the phase transition region around $T \approx T_\chi$. This indicates that the expectation values of the quark chiral condensates are nearly indistinguishable, signifying a confined state. As the system's temperature increases, the chiral condensates gradually decrease, resulting in distinct curves that characterize the nonstrange, strange, and charm chiral condensates. The smooth decrease indicates a slow transition, or {\it "crossover"}. At high temperatures above, the chiral condensates are significantly suppressed, indicating a transition to a deconfined phase in the QCD system.  The thermal evolution of the deconfinement phase transition is evident in the behavior of the Polyakov-loop order parameters  ($\phi,\; \bar{\phi}$). The order parameters of the Polyakov-loop variables are identical, i.e., $\phi = \bar{\phi}$  at $\mu_f=0$. At low temperatures (confined phase), the confined hadron phase dominates, and the order parameters of the Polyakov-loop variables vanish or take small values. As the temperature increases, the confined phase transforms smoothly to the deconfined phase, occurring at a deconfinement temperature $T_\phi$. At higher temperatures (deconfined phase), the order parameters of the Polyakov-loop variables for the Fukushima-type potential approach unity, which is related to the gluonic dynamics and the static color-charge free energy.

The middle (b) and right (c) panels of Fig.~\ref{Fig-orderParameters-T} at finite $\mu_I$, the significance of the isospin parameters ($\sigma_3$, $h_3$) becomes evident, in distinguishing between the $u-$ (dash-double-dotted curves) and $d-$ (long-dashed-dotted curves) quark condensates. During the phase transition, the normalized light quark condensates split into two separate curves representing the $u$- and $d$-quark condensates. It is apparent that the pseudo-critical temperature $T_\chi$ decreases as $\mu_I$ increases. Additionally,  the normalized strange and charm condensates remain unaffected by changes in $\mu_I$, the value of $T_\chi$ decreases as $\mu_B$ increases. Furthermore, the nonstrange and strange chiral condensates exhibit greater sensitivity to $\mu_I$ and $\mu_B$ than the charm condensate. It is noteworthy that both the isospin $\mu_I$ and baryon $\mu_B$ chemical potentials lead to the same conclusion: the phase transition occurs at a lower temperature $T_\chi$. When the isospin chemical potential $\mu_I$ is introduced, the thermal evolution of the order parameters associated with the deconfinement transition  becomes more pronounced. Specifically, the Polyakov-loop variables $\phi$ (long-dashed curves)  and its conjugate $\bar{\phi}$ (triple-dashed curves) begin to differ, i.e., $\phi \neq \bar{\phi}$. It is evident that the thermal behavior of $\phi$ consistently exceeds that of $\bar{\phi}$ as $\mu_I$ increases. Increase either $\mu_I$ or $\mu_B$ results in a gradual rise in the Polyakov-loop variables and thereby enhancing the deconfinement transition. It is significant that the effect of both $\mu_I$ and $\mu_B$ leads to the same outcome: the deconfinement transition occurs at a lower temperature $T_\phi$.

\begin{figure}[htb]
\centering{
\includegraphics[width= 16.5cm , height=5.5 cm, angle=0  ]{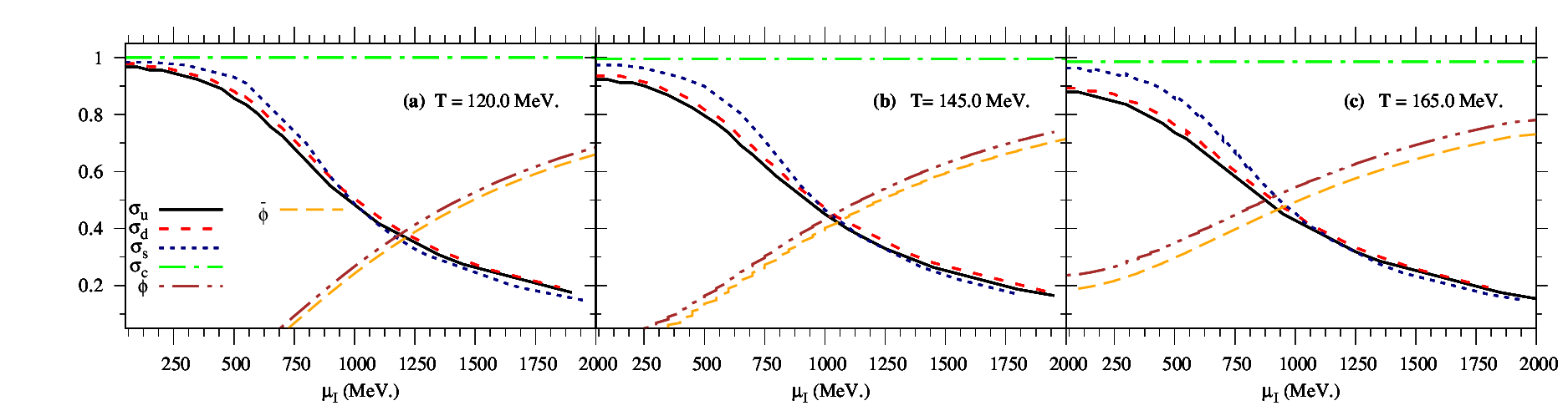}
\caption{ (Color online) The same as in Fig. \ref{Fig-orderParameters-T} but varying the isospin chemical potential at different values of temperatures $T=120.0~$MeV (left panel) $145.0~$MeV (middle panel) and $165.0~$MeV (right panel).
}
\label{Fig-orderParameters-Mu}} 
\end{figure} 
Figure~\ref{Fig-orderParameters-Mu} illustrates the behavior of the normalized chiral condensates ($\sigma_f/\sigma_{f_0}$) and the order parameters of the Polyakov-loop variables  ($\phi$ and $\bar{\phi}$) as functions of the isospin chemical potential ($\mu_I$) at different temperatures: $T = 120.0~\text{MeV}$ (left panel), $T = 145.0~\text{MeV}$ (middle panel), and $T = 165.0~\text{MeV}$ (right panel). The impact of temperature ($T$) and isospin chemical potential ($\mu_I$) on the light quark condensates ($\sigma_u$ and $\sigma_d$) and heavy quark condensates ($\sigma_s$ and $\sigma_c$) highlights significant differences between the light and heavy quark sectors. The chiral condensates ($\sigma_u$, $\sigma_d$, and $\sigma_s$) show pronounced sensitivity to both $T$ and $\mu_I$. The introduction of isospin asymmetry causes the chiral condensates of the light quark sector to become distinguishable: as $\mu_I$ increases, $\sigma_u$ decreases more rapidly than $\sigma_d$ and $\sigma_s$. In the beginning, the chiral condensates remain at their vacuum values ($\sigma_{f_0}$) at low $\mu_I$, where the system  resides in a confined phase  with chiral symmetry is spontaneously broken. As $\mu_I$ increases, the chiral condensates of the $\ell$- and $s$-quark sectors gradually decrease due to the restoration of the chiral symmetry, with a suppression near the phase transition. However, the thermal behavior of the chiral condensates exhibit a smooth phase transition or a smooth crossover. This suppression becomes more pronounced with increasing $T$. While the charm quark condensate ($\sigma_c$) is not sensitive to the isospin chemical potential, $\sigma_c$ remains nearly fixed as both $T$ and $\mu_I$ increase.
 
At lower temperatures, below the critical endpoint (CEP), i.e., $T = 120.0~\text{MeV}$ (left panel), the system remains in the confined phase, characterized by high values chiral condensates and small values of the order parameters of the Polyakov-loop variables. As $\mu_I$ increases, the transition toward the deconfined phase becomes more pronounced, with a sharper decline in chiral condensates and a significant rise in the Polyakov-loop order parameters. At temperatures above the CEP, i.e., $T = 165.0~\text{MeV}$ (right panel), the system is almost entirely in the deconfined phase. As $\mu_I$ dependence, the chiral condensates are suppressed, while the Polyakov-loop order parameters are elevated. The behavior of the order parameters of the Polyakov-loop variables ($\phi$ and $\bar{\phi}$) also exhibits isospin asymmetry. At low $\mu_I$, the two parameters are nearly equal, reflecting a symmetric QCD medium. However, as $\mu_I$ increases, $\phi$ rises more rapidly than $\bar{\phi}$. Furthermore, increasing $T$ shifts the deconfinement transition to higher values, indicating that the system transitions more rapidly from the confined to the deconfined phase. The thermal behavior of $\phi$ is more pronounced than that of its conjugate, $\bar{\phi}$. When $\mu_I$ is introduced, the evolution of the order parameters for the deconfined phase transition becomes distinguishable. It is important to highlight that both temperature ($T$) and isospin chemical potential ($\mu_I$) similarly, resulting in a lowered deconfinement transition temperature, $T_\phi$.
\begin{figure}[htb]
\centering{
\includegraphics[width= 6.0 cm, height=9.0 cm, angle=-90  ]{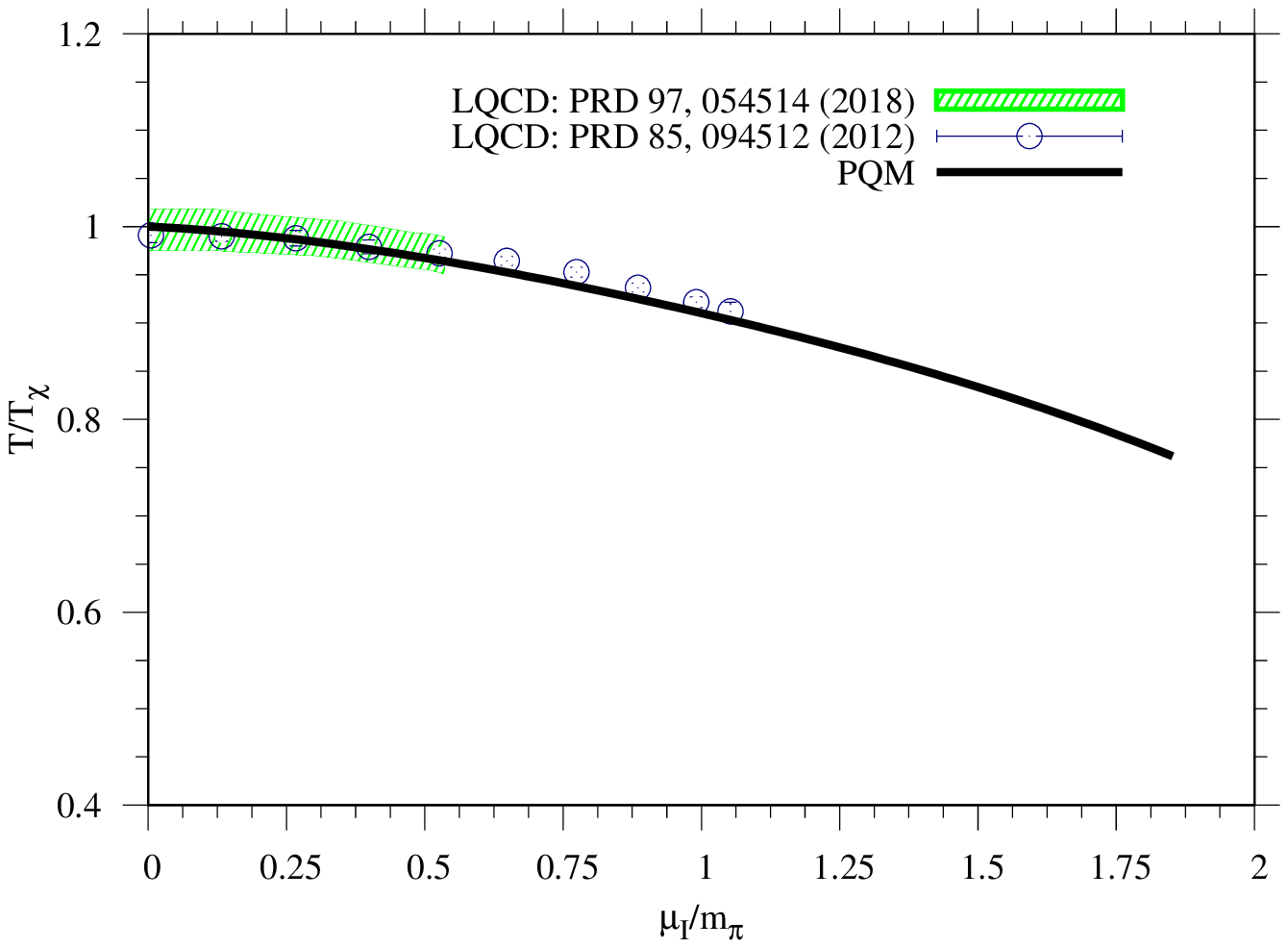}
\caption{(Color online) The chiral phase diagram ($T-\mu_I/\mu_\pi$) at vanishing baryon chemical potential. The PQM results (solid curves) are confronted with lattice QCD calculation \cite{Brandt:2017oyy} (shaded band)  and \cite{Cea:2012ev} (open symbols).
}
\label{Fig-Tmuplot}}   
\end{figure} 
The pseudo-critical temperature ($T_{\chi}$) at different values of the isospin chemical potential can be estimated by identifying the location of the inflection point as follows:
\bea
T_{\chi} =\mbox{Arg}\; \Big[ \frac{\partial \sigma_f}{\partial T} \Big]_{\mbox{max}}.
\eea  
The obtained results map out the chiral phase diagram ($T/T_{\chi}, \mu_I/m_\pi$) plane at $\mu_B = 0~$MeV. Figure~\ref{Fig-Tmuplot} illustrates the dependence of the normalized pseudo-critical temperature ($T/T_{\chi}$)  on the normalized isospin chemical potential to the pion mass ($\mu_I/m_\pi$) for the SU($4$) PQM model. The current result (solid black curve) is compared with lattice QCD calculations~\cite{Brandt:2017oyy} (shaded band) and~\cite{Cea:2012ev} (open symbols). The coordinates of ($T/T_{\chi}, \mu_I/m_\pi$) were estimated by identifying the temperature corresponding to the inflection point of the chiral susceptibility at vanishing all chemical potentials. This result reveals that the chiral phase diagram exhibits a decreasing trend in $T_{\chi}$ with increasing $\mu_I$. Furthermore, the obtained PQM curve shows good agreement with lattice QCD calculations~\cite{Cea:2012ev,Brandt:2017oyy}, particularly for $\mu_I/m_{\pi} < 1$. At higher values of $\mu_I$, where lattice QCD data is absent, the PQM model provides an extrapolation, predicting a continued suppression of the pseudo-critical temperature as $\mu_I$ increases. The PQM curve smoothly extends toward lower temperatures and higher isospin chemical potentials. This agreement validates the effectiveness of the PQM model in capturing essential QCD thermodynamic observables over a wide range of isospin chemical potentials. The SU($4$) PQM model and lattice QCD simulations have demonstrated that increasing the isospin chemical potential ($\mu_I$) leads to a decrease in the pseudo-critical temperature ($T_\chi$) associated with chiral symmetry restoration. This finding suggests that chiral symmetry is restored at lower temperatures under such conditions, potentially leading to a shift in the location of the critical-endpoint (CEP) within the QCD phase diagram. In heavy-ion collisions at NICA, the produced matter is expected to exhibit finite isospin asymmetry due to the nature of  of the colliding heavy ions, resulting in a nonzero isospin chemical potential $\mu_I$.

\subsection{QCD Equation of State and Charmed Fluctuations \label{QCDthermo}}
Thermodynamic observables play a essential role in characterising of the chiral--deconfinement phase transition for $N_f = 2+1+1$ quark flavors in a thermal and dense QCD medium. For a statistical system in equilibrium within volume $V$, the thermodynamic pressure can be estimated from the grand canonical free energy density as
$p(T, \mu_f) = \mathcal{F}_{\mathrm{PQM}}(T, \mu_f)$, and the normalized interaction measure (trace anomaly) is given by
\bea
\frac{I(T,\,\mu_f)}{T^4} =  \frac{\epsilon - 3p}{T^4} = T \; \frac{\partial}{\partial T} \Big[\frac{p}{T^4}\Big],
\eea
where the energy density is given by  $\epsilon (T,\,\mu_f)= T^2 \partial \big[p/T\big]/\partial T = -p(T,\,\mu_f) + T\,s(T,\,\mu_f)$,  while the entropy density is defined as $s(T,\,\mu_f)=\partial p/\partial T$.  In the ideal gas limit, the Stefan--Boltzmann (SB) limits can be determined from the grand canonical partition function for $(N_c^2 - 1)$ massless gluons and $N_f$ massless quark flavors, using lowest-order perturbation theory \cite{Schaefer:2007pw} as,
\bea
\frac{p_{\tiny SB}}{T^4} = (N_c^2-1) \frac{\pi^2}{45} + N_c\,N_f \Big( \frac{7\pi^2}{180} + \frac{1}{6} \Big[ \frac{\mu_f}{T}\Big]^2 + \frac{1}{12 \pi^2} \Big[ \frac{\mu_f}{T}\Big]^4\Big), \label{PSBlimit}
\eea
where the first term on the rhs of Eq.~(\ref{PSBlimit}) represents the gluonic contribution, the second term corresponds to the fermionic contribution, and the remaining terms account for the contributions of an ideal gas at finite chemical potential.
\begin{figure}[htb]
\centering{
\includegraphics[width= 6.0 cm, height=8.0 cm, angle=-90  ]{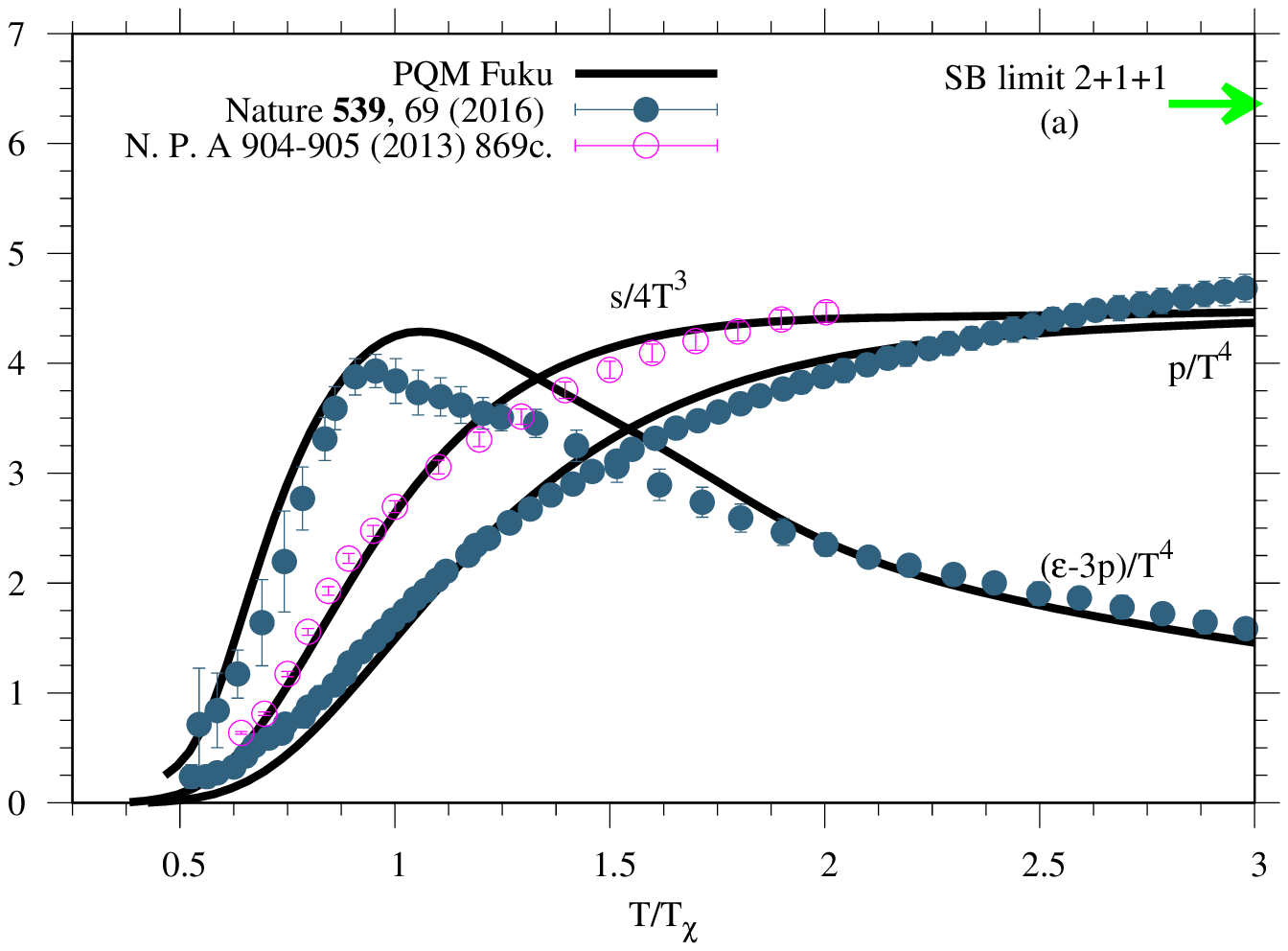}
\includegraphics[width= 6.0 cm, height=8.3 cm, angle=-90  ]{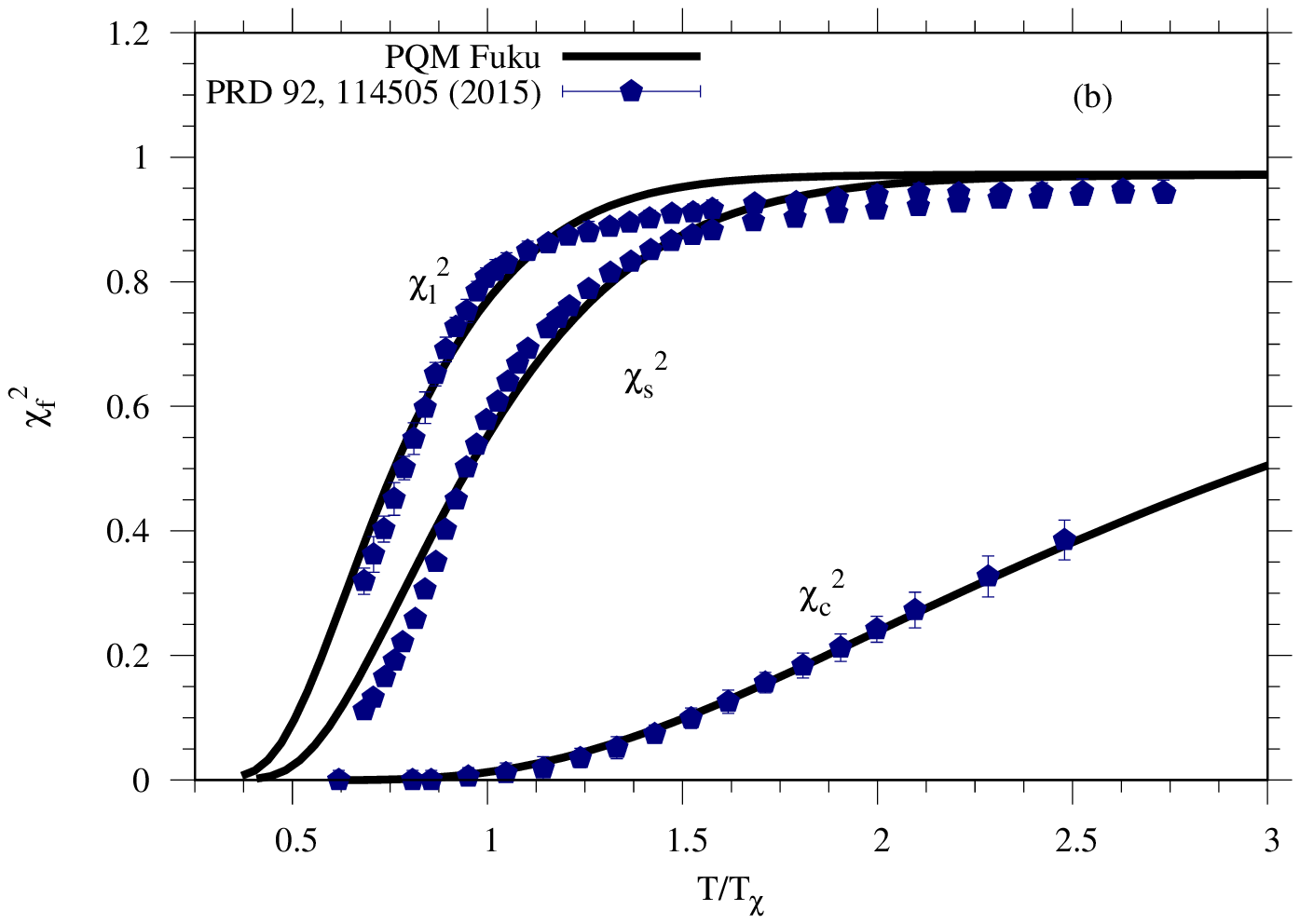}
\caption{(Color online) A comparison between the normalized PQM thermodynamic observables $p/T^4,\, I/T^4$ and $s/T^3$  (solid curves) at $\mu_f=0.0~$MeV for SU$(4)$ quark flavors with the lattice QCD simulations  \cite{Borsanyi:2016ksw} (closed circles)  and \cite{Ratti:2013uta} (open circles)  (left panel (a)). While the quark flavor susceptibilities $\chi_f^2$ are confronted with the corresponding lattice QCD calculations \cite{Bellwied:2015lba} (closed symbols) (right panel (b)).}
\label{Fig.thermo}}   
\end{figure} 
Figure~\ref{Fig.thermo} presents various dimensionless thermodynamic observables: the normalized pressure $p/T^4$, interaction measure (trace anomaly) $I/T^4$, and entropy density $s/T^3$ (left panel (a)), along with the quark flavor susceptibilities $\chi_f^2$ (right panel (b)) as functions of $T/T_\chi$ at $\mu_f = 0.0~\text{MeV}$. The results are obtained from the SU($4$) PQM model in the presence of the Polyakov-loop variables (solid curves). The PQM results show fine agreement with lattice QCD simulations~\cite{Ratti:2013uta,Borsanyi:2016ksw,Bellwied:2015lba}.

At low temperatures, the normalized pressure is suppressed due to confinement, then gradually increases with $T/T_\chi$ as new degrees of freedom emerge during the transition to the deconfined phase. At high temperatures, the pressure closes to the SB-limit, indicating that the system behaves like an massless ideal gas of free color quarks and gluons, where the interactions are effectively negligible.

 The increase in energy density with rising $T/T_\chi$ reflects the rapid liberation of degrees of freedom during the QCD phase transition, while the entropy density quantifies the disorder of the system during the hadron--quark transition. Furthermore, the interaction measure (trace anomaly) quantifies deviations from conformal symmetry, where the energy--momentum tensor $T^{\mu\nu}$ vanishes in the limit of a massless ideal gas. The trace anomaly shows a pronounced peak around $T/T_\chi \approx 1$, marking the hadron--quark phase transition. This peak corresponds to the maximal interaction effects during the crossover to the quark--gluon plasma (QGP) phase. At high temperatures, the trace anomaly decreases and approaches zero, indicating the recovery of approximate conformal symmetry.

The grand canonical ensemble treats conserved charges as external input parameters. During a heavy-ion collision, the baryon number, electric charge, and strangeness quantum numbers remain conserved. In QCD, conserved charges are associated with four quark chemical potentials that can be introduced in a $2+1+1$ quark flavor system. The set of chemical potentials $\mu_f$ is defined in Eqs.~(\ref{chemicalPota}). The higher-order susceptibilities (or fluctuations) with respect to the quark chemical potentials can be expressed in terms of derivatives of the normalized pressure in the PQM model with respect to the corresponding quark chemical potential as,
\bea
\chi_{i\,j\,k\,l}^{u\,d\,s\,c} (T,\mu_f) = \frac{\partial^{i\,j\,k\,l} \big(p(T,\mu_f) /T^4\big)}{\big(\partial\hat{\mu_u}\big)^i\,\big(\partial\hat{\mu_d}\big)^j\,\big(\partial\hat{\mu_s}\big)^k\,\big(\partial\hat{\mu_c}\big)^l},
\label{fluc_corr1}
\eea
where $\hat{\mu}_f \equiv \mu_f/T$, ensuring that the cumulants remain dimensionless, and the indices $i$, $j$, and $k$ denote the order of partial derivatives. Following the standard procedure in lattice QCD calculations, various fluctuation (diagonal) and correlation (off-diagonal)  of different conserved quantum numbers, $x = [B,\, Q,\, S,\, C]$, can be estimated from the normalized pressure in the PQM model  with respect to the corresponding chemical potential of conserved quantum number as,
\bea
\chi^{BQSC}_{\alpha\,\beta\,\gamma\, \delta}  (T,\mu_x) &=& \frac{\partial^{\alpha\,\beta\,\gamma\,\delta}\; (p(T, \hat{\mu}_x)/T^4)}{(\partial \hat{\mu}_B)^\alpha\, (\partial \hat{\mu}_Q)^\beta\, (\partial \hat{\mu}_S)^\gamma\,  (\partial \hat{\mu}_C)^\delta}, \label{fluc_corr2}
\eea
The superscripts $\alpha$, $\beta$, $\gamma$, and $\delta$ represent the orders of the derivatives, where $\hat{\mu}_x \equiv \mu_x/T$. Equations~(\ref{fluc_corr1}) and (\ref{fluc_corr2}) illustrate the (diagonal) fluctuations and (off-diagonal) correlations of quark flavors and conserved quantum numbers, respectively. The moment products or diagonal (off-diagonal) susceptibilities can be established by taking derivatives of the normalized thermodynamic pressure of the  SU($4$) PQM model with respect to the corresponding dimensionless chemical potentials, as defined in Eq.~(\ref{chemicalPota}). These moment products can be related to experimentally measured particle multiplicities, encompassing the mean, variance, skewness, kurtosis, and higher-order cumulants, as discussed in~\cite{Luo:2017faz}.

In this section, we present the quark flavor susceptibilities $\chi_f^2$ with respect to the light $\ell = \{u,\,d\}$, strange $s$, and charm $c$ quark flavors as functions of temperature $T$ at vanishing $\mu_f$. Furthermore, we show the thermal behavior of diagonal (off-diagonal) susceptibilities of the charm quantum number, as well as baryon--charm (BC) correlations and their ratios. 

The right panel of Fig.~\ref{Fig.thermo} illustrates the diagonal susceptibilities $\chi^2_\ell$, $\chi^2_s$, and $\chi^2_c$. At low temperatures, the quark flavor susceptibilities vanish since the chiral quark condensates have large values, resulting in large effective quark masses and suppressed fluctuations. At high temperatures, where the relevant degrees of freedom are liberated, quark masses become small, and temperature dominates the dynamics. The system behaves like a non-interacting ideal gas of massless quarks and gluons, and the susceptibilities close the corresponding SB-limits. The susceptibilities increase monotonically with temperature. The PQM model exhibits a smooth crossover before saturation, with values remaining below the SB-limit. Both $\chi^2_\ell$ and $\chi^2_s$ saturate at approximately $95\% $ of the SB-limit. The light quark susceptibility $\chi^2_\ell$ increases more rapidly than the strange and charm quark susceptibilities, consistent with the faster melting of the light chiral condensate. We also find that the pseudo-critical temperature associated with the charm quark susceptibility $\chi^2_c$ is higher than those for the light and strange quarks, lying beyond the crossover region. It is important to note that the PQM results incorporating the Polyakov potential, Eq.~(\ref{FukushimaPot}), improve the agreement between the PQM model and lattice QCD calculations~\cite{Bellwied:2015lba}.

The main objective here is to analyze how thermal effects on the charm quark mass are influenced by isospin asymmetry and to compare these results with available lattice QCD data. While the momentum dependence of the charm quark mass has been taken into account, the thermal behavior of $\chi^2_c$ in the low-temperature region appears fixed, with noticeable evolution only at higher temperatures. Finally, the inclusion of the charm quark susceptibility in the PQM model enhances the spatial and thermal  resolution of comparisons with lattice QCD simulations.
\begin{figure}[htb]
\centering{
\includegraphics[width= 16.5 cm, height=5.5 cm, angle=0  ]{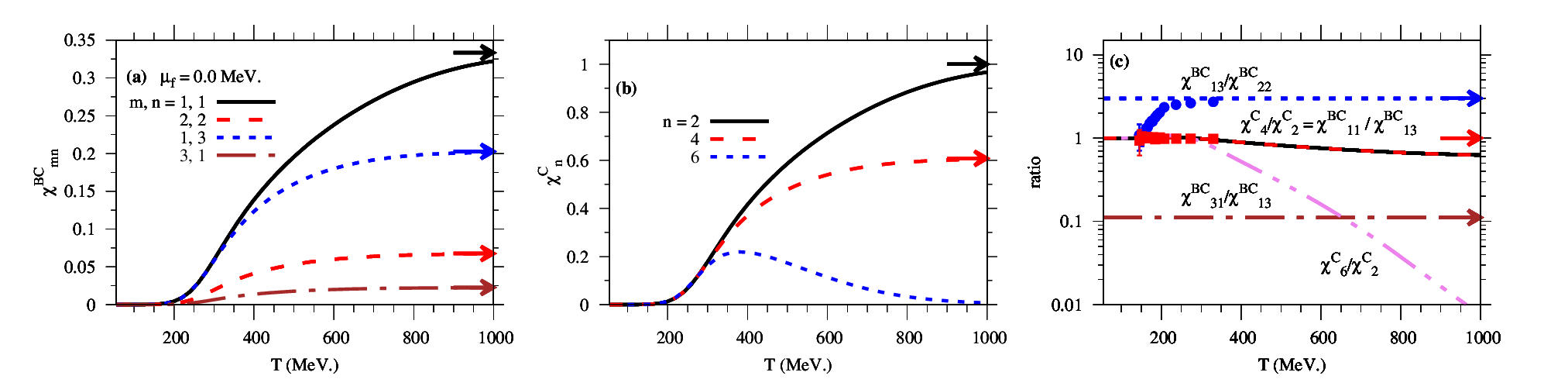}
\caption{(Color online) The thermal dependence of baryon-charm correlations (left panel (a)), higher-order charm fluctuations (middle panel (b)), and their ratios between correlations and fluctuations (right panel (c)) at vanishing chemical potential, $\mu_f=0~$MeV. The PQM results are compared with lattice QCD simulations (solid symbols) \cite{Bazavov:2014yba}. The solid arrows indicate the corresponding SB limits.}
\label{Fig. correlations}} 
\end{figure} 
Figure~\ref{Fig. correlations} shows the temperature dependence of baryon--charm correlations $\chi^{BC}_{nm}$ (left panel (a)), higher-order fluctuations of the charm quantum number $\chi^C_n$ (middle panel (b)), and the corresponding ratios between correlations and fluctuations, compared with lattice QCD simulations~\cite{Bazavov:2014yba} (right panel (c)), all at vanishing chemical potential $\mu_f = 0~\text{MeV}$. The present SU($4$) PQM results (solid curves) are obtained in the presence of the Polyakov-loop variables.

The right panel (a) of Fig.~\ref{Fig. correlations} shows the thermal behavior of the correlation between baryon and charm quantum numbers, denoted as $\chi^{BC}_{nm}$, where $(n,m)$ refers to the order of the derivatives with respect to   charm and baryon chemical potentials, respectively. The solid arrows in the plot indicate the expected SB- limits for each correlation. These correlations are sensitive to the interplay between baryon number and charm quantum number and are important observables for probing the liberation of charm degrees of freedom across the QCD crossover transition. At low temperatures, the correlations are small, indicating that charm degrees of freedom are still confined in hadronic phase. As the temperature enters the crossover region, the correlations rapidly increase,  signaling the partial restoration of chiral symmetry and the onset of deconfinement for charm-bearing states. At high temperatures, these correlations reach a saturation level below the SB-limit as indicated by arrows. Their sensitivity to charm degrees of freedom is evident, as the pseudo-critical temperature shifts lower with increasing derivative order with respect to the charm quantum number.  A systematic suppression in thermal behavior is also observed with higher-order derivatives. The influence of charm fluctuations becomes more pronounced with the inclusion of the charm chiral condensate and charm quantum number in the PQM model, enabling a more accurate evaluation of the quark--hadron phase structure at high temperatures. Additionally, the incorporation of charm meson states improves the spatial resolution of these comparisons, particularly in the crossover and high-temperature regions. 

Different $(n, m)$ combinations give a different response, where the lowest-order correlation $\chi^{BC}_{11}$ (black solid curve) rises most strongly and saturates close to its SB-limit. The higher-order derivatives such as $\chi^{BC}_{13}$ (blue dashed), $\chi^{BC}_{22}$ (red dashed) grow more slowly and saturate at lower values. While the $\chi^{BC}_{31}$ (brown dash-dot), remains suppressed over the full temperature range and contributes negligibly to the SB limit. We conclude that the sensitivity of the observables to charm fluctuations increases with the inclusion of higher derivatives, making them useful probes for testing the strength of QCD interactions, especially in the transition region relevant to heavy-ion collision experiments like those at NICA and FAIR.
  
The middle panel (b) of Fig.~\ref{Fig. correlations} illustrates the higher-order fluctuations of the charm quantum number, $\chi_n^C$, for $n = 2,\ 4,$ and $6$ in the SU($4$) PQM model. In the low-temperature region, the thermal behavior of $\chi_n^C$ shows small values, characteristic of the hadronic phase. This suppression is due to large chiral condensates, which give rise to large effective masses, thereby reducing susceptibility to fluctuations. The evolution of $\chi_n^C$ with temperature does not exhibit a sharp phase transition but rather a smooth crossover. The temperature dependence of fluctuations is significantly influenced by the chiral structure of the hadronic states. At high temperatures, a new state of matter emerges, characterized by nearly massless quarks and gluons. As the system transitions to the deconfined phase, the charm fluctuations gradually increase and approach a saturation value slightly below the SB-limit. The higher-order fluctuations, i.e., $\chi_6^C$ are more sensitive to the critical behavior of the system. Their response is sharper and peaks earlier than lower-order fluctuations.   Notably, the transition to the partonic phase becomes more rapid as the order $n$ of the derivative of the charm quantum number increases, indicating enhanced sensitivity of higher-order fluctuations to critical dynamics and deconfinement features. 

The right panel (c) of Fig.~\ref{Fig. correlations} illustrates the ratios between correlations and fluctuations of baryon--charm quantum numbers, $\chi^{BC}_{nm}/\chi^{BC}_{\alpha \beta}$, as functions of temperature $T$ at vanishing chemical potential ($\mu_f = 0~$MeV), where $n, m, \alpha, \beta > 0$ and $n + m$, $\alpha + \beta$ are even. The ratios   $\chi_4^C/\chi_2^C$, $\chi_6^C/\chi_2^C$,  $\chi^{BC}_{11}/\chi^{BC}_{13}$ and  $\chi^{BC}_{31}/\chi^{BC}_{13}$ are obtained from SU($4$) PQM model at vanishing chemical potential $\mu_f = 0~$MeV. The thermal behaviors of the ratios $\chi_4^C/\chi_2^C$ and $\chi^{BC}_{11}/\chi^{BC}_{13}$ are similar similar strength. In the low-temperature and crossover regions, these ratios begin near unity, implying that the corresponding correlations and fluctuations are comparable in strength in the hadronic phase--i.e., $\chi_4^C \approx \chi_2^C$ and $\chi^{BC}_{11} \approx \chi^{BC}_{13}$. In the hadronic phase, the PQM results agree well with lattice QCD simulations~\cite{Bazavov:2014yba}.

As the temperature increases and the system enters the crossover region, both ratios gradually decrease and then stabilize at higher temperatures, indicating that higher-order baryon--charm correlations become weaker in the deconfined phase. This behavior suggests that baryon--charm interactions follow a perturbative pattern at high $T$, with the ratios approaching values below the SB-limit. The thermal behavior of the ratio $\chi_6^C/\chi_2^C$ also begins near unity in the crossover region, confirming that charm fluctuations $\chi_2^C$ and $\chi_6^C$ are nearly equal at low temperature--representing well the contribution of charm degrees of freedom. In this region, the suppression is due to large chiral condensates, which give rise to large effective masses, leading to smaller values for higher-order susceptibilities.  As temperature increases, the ratio $\chi_6^C/\chi_2^C$ rapidly drops, reflecting a sharp change in the system near the crossover--consistent with signatures of the chiral phase transition. This decline indicates that higher-order charm fluctuations are significantly suppressed in the deconfined phase, although charm interactions remain present. The ratio $\chi^{BC}_{31}/\chi^{BC}_{13}$ and $\chi^{BC}_{13}/\chi^{BC}_{22}$ stabilizes  around $0.1$ and $3.0$, respectively. This consistent suppression indicates a strong reduction of higher-order baryon--charm correlations relative to lower-order ones and reflecting how the system evolves from a hadronic to partonic  phases.  The PQM model agrees well with lattice QCD calculations in the crossover region, confirming its reliability for studying QCD thermodynamics with charm quark effects, as it accurately captures charm-related fluctuations and matches lattice QCD results at high temperatures.
\subsection{Meson mas spectrum \label{masscurve}}
The squared masses of the meson states are defined as the second derivatives of the grand canonical  PQM thermodynamic  free energy density, Eq. (\ref{LSMPOT}), with respect to the corresponding mesonic fields at the global minimum,
\begin{equation}
m_{i,ab}^2 (T, \mu_f) = \left.\frac{\partial^2 \mathcal{F}_{\text{PQM}}(T, \mu_f)}{\partial \sigma_{i,a} \partial \sigma_{i,b}}\right|_{\text{min}} = \underbrace{m_{i,ab}^2 \big|_{\chi}}_{\text{tree-level}} + \underbrace{m_{i,ab}^2 \big|_{\text{the}}}_{\text{thermal}} + \underbrace{m_{i,ab}^2 \big|_{\text{vac}}}_{\text{vacuum}},
\label{MassMatrix}
\end{equation}
where the subscripts $\sigma_{i,a(b)}$ donates the corresponding meson field of $i-$ hadron state, $a,b\in[0,1,\cdots,15]$. Eq. (\ref{MassMatrix}) is defined as a follow:-
\begin{itemize}
\item 
The first term, $m_{i,ab}^2 \big|_{\chi}$, represents the tree-level mass matrix obtained from the second derivatives of the chiral Lagrangian with respect to the meson fields. Tables~\ref{PSTab1} and \ref{PSTab2} provide the analytical expressions for the (pseudo)scalar meson sectors, including the anomaly term in the SU($4$) PQM model. These tables  list as the follow: the first two columns assign fro the meson field in terms of $pi_i$ Pseudoscalar meson field and  PDG counterparts~\cite{ParticleDataGroup:2022pth}. While the third column given the quark  flavor content based on SU($4$) symmetry. Furthermore  the tree-level mass-squared derived analytically from the SU($4$) PQM chiral Lagrangian, Eq.(\ref{ChrLAge}).  

In Tab.~\ref{PSTab1}, the squared mass of the $K$ meson state corresponds to the ($4,4$) element of the pseudoscalar mass matrix and satisfies the Gell-Mann--Okubo mass relation~\cite{Okubo:1961jc,gell1966eightfold,Burakovsky:1997sd}. The $\eta$ and $\eta^\prime$ masses result from the diagonalization of the isoscalar pseudoscalar mass matrix, which incorporates off-diagonal flavor mixing and anomaly terms~\cite{Schaefer:2008hk}. The charm isosinglet $\eta_c$ includes $\sigma_c$ contributions and is further shifted due to charm-sector symmetry breaking. The $D$ resonances include charm-specific symmetry-breaking terms ($\epsilon_c$). A similar approach applies to Tab.~\ref{PSTab2}, but for scalar meson states. The $\kappa$ meson state corresponds to the ($4,4$) element of the scalar mass matrix and also satisfies the Gell-Mann--Okubo mass relation~\cite{Okubo:1961jc,gell1966eightfold,Burakovsky:1997sd}. The isoscalar states $f_0$ and $\sigma$ in the scalar multiplet are obtained via the diagonalization of the isoscalar scalar mass matrix, which includes off-diagonal flavor mixing and anomaly terms, as presented in Tab.~\ref{PSTab2}~\cite{Schaefer:2008hk}.
\end{itemize}

{\small
\begin{table}[htb]
\begin{tabular}{ l|c|c|c|p{6.0cm} }
\toprule
\multicolumn{1}{c|}{\bf Field}   & {\bf PDG} (MeV)  & \begin{tabular}[c]{@{}c@{}} {\bf Quark}\\{\bf content} \end{tabular}      & \begin{tabular}[c]{@{}c@{}}{\bf Mass}\\{\bf squared}\end{tabular}                                                                 & \multicolumn{1}{|c}{\bf Analytical derivation} \\ 
\midrule \midrule
\begin{tabular}[c]{@{}l@{}}$\pi^\pm=(\pi_1\pm i \pi_2)/\sqrt{2} $,\\  $\pi^0=\pi_3$\end{tabular}                                                                 & $\pi(1300)$ & \begin{tabular}[c]{@{}c@{}}$u\bar{d}, d\bar{u}$ \\ $ \frac{u\bar{u}-d\bar{d}}{\sqrt{2}}$\end{tabular}    & $m_{\pi}^2 $ & $ m_0^2 +\big(\lambda_1 +\frac{\lambda_2}{2} \big)  \big( \frac{\sigma_u^2}{2} + \frac{\sigma_d^2}{2}\big) + \lambda_1 \big( \sigma_s^2 +\sigma_c^2\big) -   \frac{\mathcal{C}}{2}  \sigma_s  \sigma_c$.                       \\ \hline
\begin{tabular}[c]{@{}l@{}}$K^\pm=(\pi_4\pm i \pi_5)/\sqrt{2}$,\\ $K^0 =(\pi_6+ i \pi_7)/\sqrt{2} $, \\$\overline{K}^0=(\pi_6 - i \pi_7)/\sqrt{2}$\end{tabular} &$K(1460)$  & \begin{tabular}[c]{@{}c@{}c@{}}$u\bar{s}, s\bar{u}$ \\ $d\bar{s}$\\ $s\bar{d}$ \end{tabular}    & $m^2_K$  & $ m_0^2 + \big(\lambda_1 +\frac{\lambda_2}{2}\big) \big( \frac{\sigma_u^2}{2} + \frac{\sigma_d^2}{2} \big) -  \lambda_2   \big( \frac{\sigma_u}{\sqrt{2}} + \frac{\sigma_d}{\sqrt{2}} \big) \sigma_s+\big(\lambda_1 + \lambda_2\big) \sigma^2_s+ \lambda_1 \sigma_c^2  -  \frac{\mathcal{C}}{2}  \sigma_s  \sigma_c  $. \\ \hline
$\eta_\ell$ &$\eta_\ell (1408)$&   $ \frac{u\bar{u}+d\bar{d}}{\sqrt{2}}$ & $m_{\eta_\ell}^2$ & $ m_0^2 +\big(\lambda_1 +\frac{\lambda_2}{2} \big)  \big( \frac{\sigma_u^2}{2} + \frac{\sigma_d^2}{2}\big) + \lambda_1 \big( \sigma_s^2 +\sigma_c^2\big) +   \frac{\mathcal{C}}{2}  \sigma_s  \sigma_c  $.\\ \hline
$\eta_s$ &$\eta_s(1475)$& $s\bar{s}$  & $m_{\eta_s}^2$  &$m_0^2 +\lambda_1 \big( \frac{\sigma_u^2}{2} + \frac{\sigma_d^2}{2} + \sigma_s^2 +\sigma_c^2\big) + \lambda_2 \sigma_s^2$.\\ \hline
 Eigenv$^-$ ($\eta_\ell,\eta_s$)& $\eta (1294)$& \begin{tabular}[c]{@{}c@{}} $  \frac{u\bar{u}-d\bar{d}}{\sqrt{2}}$ \\ $s\bar{s}$ \end{tabular}&$m_{\eta}^2$ &  $\frac{1}{2} \Big(m_{\eta_\ell}^2 +m_{\eta_s}^2 - \Delta_{\eta_\ell,\eta_s} \Big)$ .\\ \hline
 Eigenv$^+$ ($\eta_\ell,\eta_s$) &$\eta^\prime (958)$ &  \begin{tabular}[c]{@{}c@{}} $  \frac{u\bar{u}-d\bar{d}}{\sqrt{2}}$ \\ $s\bar{s}$ \end{tabular} & $m_{\eta^\prime}^2$  & \begin{tabular}[c]{@{}l@{}}$\frac{1}{2} \Big(m_{\eta_\ell}^2 +m_{\eta_s}^2 + \Delta_{\eta_\ell,\eta_s} \Big)$,\\ $\Delta_{\eta_\ell,\eta_s} =\sqrt{\big(m_{\eta_\ell}^2 - m_{\eta_s}^2 \big)^2 + 4m_{{\eta_\ell,\eta_s}}^2 }$, \\$ m_{\eta_\ell,\eta_s}^2 =   \frac{\mathcal{C}}{2}   \big( \frac{\sigma_u}{2} + \frac{\sigma_d}{2}\big) \sigma_c$.\end{tabular}\\ \hline
$\eta_c=(\pi_0-\sqrt{3} \pi_{15})/2  $ & $\eta_c (2983)$  & $c\bar{c}$  &  $m_{\eta_c}^2 $ & $  m_0^2 +\lambda_1 \big( \frac{\sigma_u^2}{2} + \frac{\sigma_d^2}{2} + \sigma_s^2 +\sigma_c^2\big)  + \lambda_2 \sigma_c^2 +2 \varepsilon_c$. \\ \hline
\begin{tabular}[c]{@{}l@{}}$D_0^\pm =(\pi_{11} \pm i \pi_{12})/\sqrt{2}$,\\ $D_0^0 =(\pi_{9} - i \pi_{10})/\sqrt{2} $, \\$\overline{D}_0^0=(\pi_{9} + i \pi_{10})/\sqrt{2}$\end{tabular} & $D^0(1865)$ & $u\bar{c},c\bar{u}$  &  $m_{D}^2$ &$  m_0^2 + \big(\lambda_1 +\frac{\lambda_2}{2} \big)  \big( \frac{\sigma_u^2}{2} + \frac{\sigma_d^2}{2}\big) + \lambda_1 (\sigma_s^2 + \sigma_c^2) + \lambda_2 \sigma_c^2 -\frac{\lambda_2}{\sqrt{2}} \big( \frac{\sigma_u}{2} + \frac{\sigma_d}{2}\big) \sigma_c + \varepsilon_c  $. \\ \hline
$D_{S,0}^\pm=(\pi_{13} \pm i \pi_{14})/\sqrt{2}$ & $D_{S}^\pm (1968)$ & $s\bar{c}, c\bar{s}$ &   $m_{D_s}^2 $ &  $  m_0^2 + \lambda_1 \big( \frac{\sigma_u^2}{2} + \frac{\sigma_d^2}{2} + \sigma_s^2 +\sigma_c^2\big) +\lambda_2 (\sigma_s^2 + \sigma_c^2) -\lambda_2 \sigma_s \sigma_c +  \varepsilon_c $. \\ \hline
\end{tabular}
\caption{Mass expression of spin $J^{\mathcal{P} \mathcal{C}} = 0^{-+}$ Pseudoscalar meson sectors, PDG correspondence, quark content, quantum numbers and PDG mass values from \cite{ParticleDataGroup:2022pth}.  \label{PSTab1}}
\end{table}
}

{\small  
\begin{table}[htb]
\begin{tabular}{ l|c|c|c|p{6.0cm} }
\toprule
\multicolumn{1}{c|}{\bf Field}                                                                                                                                      & {\bf PDG} (MeV)  & \begin{tabular}[c]{@{}c@{}} {\bf Quark}\\{\bf content} \end{tabular}                                                                   & \begin{tabular}[c]{@{}c@{}}{\bf Mass}\\{\bf squared}\end{tabular}                                                                 & \multicolumn{1}{|c}{\bf Analytical derivation} \\ 
\midrule \midrule
\begin{tabular}[c]{@{}l@{}}$a_0^\pm=(\sigma_1\pm i \sigma_2)/\sqrt{2} $,\\  $a_0^0=\sigma_3$\end{tabular}                                                                 & $a_0(1474)$ & \begin{tabular}[c]{@{}c@{}}$u\bar{d}, d\bar{u}$ \\ $ \frac{u\bar{u}-d\bar{d}}{\sqrt{2}}$\end{tabular} &   $m_{a_0}^2 $ & $ m_0^2 +\lambda_1 \big(\sigma_u^2 +\sigma_d^2 - \sigma_u\sigma_d + \sigma_s^2 +\sigma_c^2\big) + \frac{3\lambda_2}{2}\big( \frac{\sigma_u^2}{2} + \frac{\sigma_d^2}{2}\big) +\frac{\mathcal{C}}{2}  \sigma_s  \sigma_c$,                     \\ \hline
\begin{tabular}[c]{@{}l@{}}$\kappa^\pm=(\sigma_4\pm i \sigma_5)/\sqrt{2}$,\\ $\kappa^0 =(\sigma_6+ i \sigma_7)/\sqrt{2} $, \\$\overline{\kappa}^0=(\sigma_6 - i \sigma_7)/\sqrt{2}$\end{tabular} &$k^*_0(1425)$  & \begin{tabular}[c]{@{}c@{}c@{}}$u\bar{s}, s\bar{u}$ \\ $d\bar{s}$\\ $s\bar{d}$ \end{tabular} &   $m^2_\kappa$  & $  m_0^2 + \big(\lambda_1 +\frac{\lambda_2}{2}\big) \big( \frac{\sigma_u^2}{2} + \frac{\sigma_d^2}{2} \big)  +\lambda_2   \big( \frac{\sigma_u}{\sqrt{2}} + \frac{\sigma_d}{\sqrt{2}} \big) \sigma_s+\big(\lambda_1 + \lambda_2\big) \sigma^2_s+ \lambda_1 \sigma_c^2 + \frac{\mathcal{C}}{2}  \sigma_s  \sigma_c  $ \\ \hline
$\sigma_\ell$ &$\sigma_\ell (1370)$& $ \frac{u\bar{u}+d\bar{d}}{\sqrt{2}}$& $m_{\sigma_\ell}^2$ &  $m_0^2 +\lambda_1 \big(\sigma_u^2 +\sigma_d^2 + \sigma_u\sigma_d + \sigma_s^2 +\sigma_c^2\big) + \frac{3\lambda_2}{2}\big( \frac{\sigma_u^2}{2} + \frac{\sigma_d^2}{2}\big) -\frac{\mathcal{C}}{2}  \sigma_s  \sigma_c$,\\ \hline
$\sigma_s$ &$\sigma_s(1506)$& $s\bar{s}$&  $m_{\sigma_s}^2$  &$m_0^2 + \lambda_1\big( \frac{\sigma_u^2}{2} + \frac{\sigma_d^2}{2}\big) + 3(\lambda_1+\lambda_2)\sigma s^2 + \lambda_1 \sigma_c^2,$\\ \hline
 Eigenv$^-$ ($\sigma_\ell,\sigma_s$)& $f_0(1370)$& \begin{tabular}[c]{@{}c@{}} $ \frac{u\bar{u}+d\bar{d}}{\sqrt{2}}$ \\ $s\bar{s}$ \end{tabular} & $m_{f_0}^2$ &  $\frac{1}{2} \Big(m_{\sigma_\ell}^2 +m_{\sigma_s}^2 - \Delta_{\sigma_\ell,\sigma_s} \Big)$ .\\ \hline
 Eigenv$^+$ ($\sigma_\ell,\sigma_s$) &$\sigma (1200)$ &  \begin{tabular}[c]{@{}c@{}} $ \frac{u\bar{u}+d\bar{d}}{\sqrt{2}}$ \\ $s\bar{s}$ \end{tabular} &  $m_{\sigma}^2$  & \begin{tabular}[c]{@{}l@{}}$\frac{1}{2} \Big(m_{\sigma_\ell}^2 +m_{\sigma_s}^2+ \Delta_{\sigma_\ell,\sigma_s} \Big)$,\\ $\Delta_{\sigma_\ell,\sigma_s} =\sqrt{\big(m_{\sigma_\ell}^2 - m_{\sigma_s}^2 \big)^2 + 4m_{{\sigma_\ell,\sigma_s}}^2 }$, \\$ m_{\sigma_\ell,\sigma_s}^2 = \lambda_1  \big( \frac{\sigma_u}{2} + \frac{\sigma_d}{2}\big) \sigma_s - \frac{\mathcal{C}}{2}   \big( \frac{\sigma_u}{2} + \frac{\sigma_d}{2}\big)\sigma_c$.\end{tabular} .\\ \hline
$\chi_{c_0}=(\sigma_3-\sqrt{3} \pi_{15})/2  $ & $\chi_{c_0} (3415)$  & $c\bar{c}$ &   $m_{\chi_{c_0}}^2 $ & $ m_0^2 +\lambda_1 \big( \frac{\sigma_u^2}{2} + \frac{\sigma_d^2}{2} + \sigma_s^2 \big)  + 3 (\lambda_1+ \lambda_2) \sigma_c^2 +2 \varepsilon_c$, \\ \hline
\begin{tabular}[c]{@{}l@{}}$D_0^\pm =(\sigma_{11} \pm i \sigma_{12})/\sqrt{2}$,\\ $D_0^0 =(\sigma_{9} - i \sigma_{10})/\sqrt{2} $, \\$\overline{D}_0^0=(\sigma_{9} + i \sigma_{10})/\sqrt{2}$\end{tabular} & $D_0^*(2350)$ & $u\bar{c},c\bar{u}$  &   $m_{D_0^*}^2$ &$  m_0^2 + \big(\lambda_1 +\frac{\lambda_2}{2} \big)  \big( \frac{\sigma_u^2}{2} + \frac{\sigma_d^2}{2}\big) + \lambda_1 (\sigma_s^2 + \sigma_c^2) + \lambda_2 \sigma_c^2 +\frac{\lambda_2}{\sqrt{2}} \big( \frac{\sigma_u}{2} + \frac{\sigma_d}{2}\big) \sigma_c + \varepsilon_c  $. \\ \hline
$D_{S,0}^\pm=(\sigma_{13} \pm i \sigma_{14})/\sqrt{2}$ & $D_{s0}*(2318)$ & $s\bar{c}, c\bar{s}$ &   $m_{D_{s0}*}^2 $ & $ m_0^2 + \lambda_1 \big( \frac{\sigma_u^2}{2} + \frac{\sigma_d^2}{2} + \sigma_s^2 +\sigma_c^2\big) +\lambda_2 (\sigma_s^2 + \sigma_c^2) +
\lambda_2 \sigma_s \sigma_c +  \varepsilon_c  $. \\ \hline
\end{tabular}
\caption{ The same as in Tab. \ref{PSTab1} but for spin $ J^{\mathcal{P} \mathcal{C}} = 0^{++}$ scalar meson sectors. \label{PSTab2}}
\end{table}
}
\begin{itemize}
\item The second term $m_{i,ab}^2 \big|_{\text{the}}$ arises from the fermionic (partonic) contribution to the grand thermodynamic free energy, which describes the thermal and dense QCD-medium effects on meson sectors due to the quark--antiquark contribution, as given in Eq.~(\ref{qaurkPot}) within the SU($4$) PQM model~\cite{Schaefer:2008hk,Tawfik:2014gga}.
\bea
m_{i,ab}^2 \big|_{\text{the}} &=& \left.  \mathcal{N}_c \sum_{f} \int_0^{\infty} \frac{d^3p}{(2\pi)^3}  \right.  \frac{1}{2E_{q,f}} \biggl[ (\mathcal{N}_{q_f} + \mathcal{N}_{\bar{q}_f} ) 
\biggl( m^{2}_{f,ab} - \frac{m^{2}_{f,a} m^{2}_{f,b}}{2 E_{q,f}^{2}} \biggr) \nn \\
&\quad +& (\mathcal{B}_{q_f} + \mathcal{B}_{\bar{q}_f}) \biggl(\frac{m^{2}_{f,a}  m^{2}_{f,b}}{2 E_{q,f} T} \biggr) \biggr], \label{eq:ftmass}
\eea

where $\mathcal{N}_c = 2\,N_c$ color flavors. This expression includes the squared quark masses with respect to meson fields from the SU($4$) PQM model, Eq.~(\ref{eq:ftmass}). The diagonalizations of the quark-mass matrix are listed in Tab.~\ref{Smesonfields}. The quark mass derivative with respect to the meson fields $\sigma_{i,a}$ is defined as $m^2_{f,a} \equiv \partial m^2_f/\partial \sigma_{i,a}$, and the second derivative with respect to $\sigma_{i,a}$ and $\sigma_{i,b}$ is given by $m^2_{f,{ab}} \equiv \partial^2 m^2_f/\partial \sigma_{i,a} \partial \sigma_{i,b}$. More details can be found in Refs.~\cite{Gupta:2009fg,Tawfik:2014gga}. Tab.~\ref{Smesonfields} lists the squared quark masses in the SU($4$) PQM model with respect to the meson fields: the first four columns correspond to the light ($u$ and $d$) quarks, the fifth and sixth columns to the strange ($s$) quark, and the seventh and eighth columns to the charm ($c$) quark. The terms $E_{q,f}(T,\mu)$ are symmetric under $\mu \rightarrow -\mu$, i.e., $E_{q,f}(T,\mu) = E_{q,f}(T,-\mu)$, and thus ensure that the thermodynamic contributions from quark--antiquark pairs remain even functions of the chemical potential. This symmetry plays an essential role in conserving CP invariance and simplifies the evaluation of thermodynamic integrals at vanishing and finite $\mu$ as,
\bea
\mathcal{N}_{q_f}  &=& \frac{\phi e^{-\,E_{q,f}/T} + 2 \bar{\phi} e^{- 2\, E_{q,f}/T} + e^{-3\,E_{q,f}/T}}{1+3(\phi+\bar{\phi} e^{-E_{q,f}/T}) e^{-E_{q,f}/T}+e^{-3E_{\bar{q},f}/T}}, \\
 \mathcal{N}_{\bar{q}_f} &=& \frac{\bar{\phi} e^{-E_{\bar{q},f}/T} + 2 \phi e^{- 2 E_{\bar{q},f}/T} + e^{-3E_{\bar{q},f}/T}}{ 1+3(\bar{\phi}+\phi e^{-E_{\bar{q},f}/T}) e^{-E_{\bar{q},f}/T}+e^{-3E_{\bar{q},f}/T}},
\eea
are implemented as in Ref. \cite{Gupta:2009fg}. Furthermore, for (anti-)quark $\mathcal{B}_{q_f (\bar{q}_f)} =3 \mathcal{N}_{q_f (\bar{q}_f)}^2 - \mathcal{C}_{q_f (\bar{q}_f)}$ where
\bea
\mathcal{C}_{{q}_f} &=& \frac{\phi e^{-\,E_{q,f}/T} +4 \bar{\phi} e^{-2\,E_{q,f}/T} +3 e^{-3\,E_{q,f}/T}}{1+3(\phi+\bar{\phi} e^{-E_{q,f}/T})\, e^{-E_{q,f}/T}+e^{-3E_{\bar{q},f}/T}}, \\
\mathcal{C}_{\bar{q}_f} &=&  \frac{\bar{\phi} e^{-E_{\bar{q},f}/T} + 4 \phi e^{-2E_{\bar{q},f}/T} +3 e^{-3E_{\bar{q},f}/T}}{ 1+3(\bar{\phi}+\phi e^{-E_{\bar{q},f}/T})\, e^{-E_{\bar{q},f}/T}+e^{-3E_{\bar{q},f}/T}},
\eea
are defined as in Ref.  \cite{Gupta:2009fg}. 
\end{itemize}

\begin{table}[h!]
\centering
\begin{tabular}{l|cc|cc|cc|cc}
\toprule
 { Field} & {\bf  $\frac{m_{u,\alpha}^2 m_{u,\beta}^2}{g^4}$ }& {\bf$\frac{m_{u,\alpha\beta}^2 }{g^2}$  }&{\bf  $\frac{m_{d,\alpha}^2 m_{d,\beta}^2}{g^4}$ } &{\bf $\frac{m_{d,\alpha\beta}^2}{g^2}$}  &{\bf $\frac{m_{s,\alpha}^2 m_{s,\beta}^2}{g^4}$ }& {\bf $\frac{m_{s,\alpha\beta}^2}{g^2}$} &{\bf $\frac{m_{c,\alpha}^2 m_{c,\beta}^2 }{g^4} $}  & {\bf $\frac{m_{c,\alpha\beta}^2}{g^2}$ } \\
\midrule
$\sigma_{0} \sigma_{0}$ & 
$\frac{1}{8} \sigma_u^2$  &    $\frac{1}{4}$&      
$\frac{1}{8} \sigma_d^2$  &   $\frac{1}{4}$ &
$\frac{1}{4} \sigma_s^2$ & $\frac{1}{4}$ & 
$\frac{1}{4} \sigma_c^2$ &  $\frac{1}{4}$ \\
$\sigma_0  \sigma_3$ &
$\frac{\sigma_u^2}{4\sqrt{2}} $  &  $\frac{1}{2\sqrt{2}}$ &
$-\frac{\sigma_d^2}{4\sqrt{2}}$ & $-  \frac{1}{2\sqrt{2}}$ &
$0$ & $0$  &
$0$ & $0$  \\

$\sigma_{0} \sigma_{8}$ & 
$\frac{1}{4\sqrt{6}} \sigma_u^2$ & $\frac{1}{2\sqrt{6}}$ &
$\frac{1}{4\sqrt{6}} \sigma_d^2$&  $\frac{1}{2\sqrt{6}}$ &
$-\frac{\sqrt{2}}{3} \sigma_s^2$ & $\frac{1}{\sqrt{6}}$ &
$0$ & $0$ \\

$\sigma_0  \sigma_{15}$ &
$\frac{1}{8\sqrt{3}}\sigma_u^2$ &  $\frac{1}{4\sqrt{3}}$ &
$\frac{1}{8\sqrt{3}}\sigma_d^2$ &  $\frac{1}{4\sqrt{3}}$ &
$-\frac{1}{4\sqrt{3}}\sigma_s^2$ & $\frac{1}{4\sqrt{3}}$ &
$\frac{\sqrt{3}}{4} \sigma_c^2$ & $-\frac{\sqrt{3}}{4}$ \\
$\sigma_{3} \sigma_{3}$ &
$\frac{1}{4} \sigma_u^2$ & $\frac{1}{2}$ &
$-\frac{1}{4} \sigma_d^2$ & $-\frac{1}{2}$ &
$0$ & $0$ &
$0$ & $0$ \\
$\sigma_3   \sigma_8$ &
$\frac{1}{4\sqrt{3}} \sigma_u^2$& $\frac{1}{2\sqrt{3}}$ &
$-\frac{1}{4\sqrt{3}} \sigma_u^2$ & $-\frac{1}{2\sqrt{3}}$&
0 & 0 &
0 & 0 \\
$\sigma_3   \sigma_{15}$ &
$\frac{1}{4\sqrt{6}} \sigma_u^2$& $\frac{1}{2\sqrt{6}}$ &
$-\frac{1}{4\sqrt{6}} \sigma_u^2$ & $-\frac{1}{2\sqrt{6}}$&
0 & 0 &
0 & 0 \\
$\sigma_{8} \sigma_{8}$ &
$\frac{1}{12} \sigma_u^2$ & $\frac{1}{6}$ &
$\frac{1}{12} \sigma_d^2$ & $\frac{1}{6}$ & 
$\frac{2}{3} \sigma_s^2$ & $\frac{2}{3}$ & 
$0$ & $0$ \\
$\sigma_8  \sigma_{15}$ &
$\frac{1}{12\sqrt{2}} \sigma_u^2$ & $\frac{\sqrt{1}}{6\sqrt{2}}$ &
$\frac{1}{12\sqrt{2}} \sigma_d^2$ & $\frac{\sqrt{1}}{6\sqrt{2}}$ &
$\frac{1}{3\sqrt{2}} \sigma_s^2$ & $\frac{\sqrt{1}}{3\sqrt{2}}$ &
0 & 0 \\
$\sigma_{15} \sigma_{15}$ & 
$\frac{1}{24} \sigma_u^2$ & $\frac{1}{12}$ &
$\frac{1}{24} \sigma_d^2$ & $\frac{1}{12}$ & 
$\frac{1}{12} \sigma_s^2$ & $\frac{1}{12}$ & 
$\frac{3}{4} \sigma_c^2$ & $\frac{3}{4}$ \\

$\sigma_{1} \sigma_{1}$ & 
$\sigma_u^2$ & $1$ & 
$\sigma_d^2$ & $1$ & $0$ & $0$ & $0$ & $0$ \\

$\sigma_{4} \sigma_{4}$ &
$0$ & $ \sigma_u \frac{\sigma_u + \sqrt{2} \sigma_s}{\sigma_u^2 - 2 \sigma_s^2}$ &
$0$ & $\sigma_d \frac{\sigma_d + \sqrt{2} \sigma_s}{\sigma_d^2 - 2 \sigma_s^2}$ &
$0$ & $ \sigma_s \frac{2\sigma_s + \sqrt{2} \sigma_{\ell}}{2 \sigma_s^2 - \sigma_{\ell}^2}$ & $0$ & $0$ \\

$\sigma_{6} \sigma_{6}$ &
0&0&
0&0&
0& $ \sigma_s \frac{\sigma_s + \sqrt{3} \sigma_{c}}{\sigma_s^2 -3\sigma_{c}^2}$&
0& $ \sigma_c \frac{\sqrt{3} \sigma_s + 3 \sigma_{c}}{3\sigma_c^2 -\sigma_{s}^2}$ \\ 

$\pi_{0} \pi_{0}$ & 
$0$   &    $\frac{1}{4}$&      
$0$   &   $\frac{1}{4}$ &
$\frac{1}{4} \sigma_s^2$ & $\frac{1}{4}$ & 
$\frac{1}{4} \sigma_c^2$ &  $\frac{1}{4}$ \\
$\pi_0  \pi_3$ &
$0$   &  $\frac{1}{2\sqrt{2}}$ &
$0$  & $-  \frac{1}{2\sqrt{2}}$ &
$0$ & $0$  &
$0$ & $0$  \\

$\pi_{0} \pi_{8}$ & 
$0$  & $\frac{1}{2\sqrt{6}}$ &
$0$ &  $\frac{1}{2\sqrt{6}}$ &
$-\frac{\sqrt{2}}{3} \sigma_s^2$ & $\frac{1}{\sqrt{6}}$ &
$0$ & $0$ \\

$\pi_0  \pi_{15}$ &
$0$  &  $\frac{1}{4\sqrt{3}}$ &
$0$  &  $\frac{1}{4\sqrt{3}}$ &
$-\frac{1}{4\sqrt{3}}\sigma_s^2$ & $\frac{1}{4\sqrt{3}}$ &
$\frac{\sqrt{3}}{4} \sigma_c^2$ & $-\frac{\sqrt{3}}{4}$ \\
$\pi_{3}\pi_{3}$ &
$0$ & $\frac{1}{2}$ &
$0$  & $-\frac{1}{2}$ &
$0$ & $0$ &
$0$ & $0$ \\
$\pi_3   \pi_8$ &
$0$ & $\frac{1}{2\sqrt{3}}$ &
$0$  & $-\frac{1}{2\sqrt{3}}$&
0 & 0 &
0 & 0 \\
$\pi_3   \pi_{15}$ &
$0$ & $\frac{1}{2\sqrt{6}}$ &
$0$  & $-\frac{1}{2\sqrt{6}}$&
0 & 0 &
0 & 0 \\
$\pi_{8} \pi_{8}$ &
$0$  & $\frac{1}{6}$ &
$0$  & $\frac{1}{6}$ & 
$0$  & $\frac{2}{3}$ & 
$0$ & $0$ \\
$\pi_8  \pi_{15}$ &
$0$  & $\frac{\sqrt{1}}{6\sqrt{2}}$ &
$0$  & $\frac{\sqrt{1}}{6\sqrt{2}}$ &
$0$  & $\frac{\sqrt{1}}{3\sqrt{2}}$ &
0 & 0 \\
$\pi_{15} \pi_{15}$ & 
$0$  & $\frac{1}{12}$ &
$0$  & $\frac{1}{12}$ & 
$0$  & $\frac{1}{12}$ & 
$0$  & $\frac{3}{4}$ \\

$\pi_{1} \pi_{1}$ & 
$0$ & $1$ & 
$0$ & $1$ & $0$ & $0$ & $0$ & $0$ \\

$\pi_{4} \pi_{4}$ &
$0$ & $ \sigma_u \frac{\sigma_u - \sqrt{2} \sigma_s}{\sigma_u^2 - 2 \sigma_s^2}$ &
$0$ & $\sigma_d \frac{\sigma_d - \sqrt{2} \sigma_s}{\sigma_d^2 - 2 \sigma_s^2}$ &
$0$ & $ \sigma_s \frac{2\sigma_s - \sqrt{2} \sigma_{\ell}}{2 \sigma_s^2 - \sigma_{\ell}^2}$ & $0$ & $0$ \\

$\pi_{6} \pi_{6}$ &
0&0&
0&0&
0& $ \sigma_s \frac{\sigma_s - \sqrt{3} \sigma_{c}}{\sigma_s^2 -3\sigma_{c}^2}$&
0& $ \sigma_c \frac{\sqrt{3} \sigma_s + 3 \sigma_{c}}{3\sigma_c^2 -\sigma_{s}^2}$ \\ 

\hline
\end{tabular}
\caption{Squared--quark mass second derivatives with respect to the relevant meson fields within SU($4$) PQM Model. \label{Smesonfields}}
\end{table}
  
\begin{itemize}
\item The third term  $m_{i,ab}^2 \big|_{\text{vac}}$ defines the fermionic vacuum correction \cite{Chatterjee:2011jd, Kovacs:2016juc}, using the  three dimensional--cutoff $\Lambda$ in fermionic vacuum term in Eg. \ref{qaurkPot}, one obtains
\bea
m_{i,ab}^2  \Bigg\vert_{\mbox{vac}} &=& \left. -\frac{\mathcal{N}_c}{(4\pi)^2} \sum_{f}  \right. \Big[ \Big( \ln{\big[\frac{m_f}{\Lambda}\big]} +\frac{3}{2}\Big) m^2_{f,a} m^2_{f,b} + \frac{m_f^2}{2}\Big(1 + 4 \ln{\big[\frac{m_f}{\Lambda}\big]}   \Big) m^2_{f,{ab}} \Big].
\eea
\end{itemize}
The present calculations of meson state squared masses account for both the isospin asymmetry of light quarks ($m_u \neq m_d$) and the inclusion of heavy quark masses ($m_s$ and $m_c$). Tab.~\ref{tab:meson masses} provides a comprehensive comparison of the (pseudo)scalar meson masses in the SU($3$) (third column) and SU($4$) (sixth column) formulations of the PQM model, both incorporating isospin asymmetry. Our results are compared with other effective thermal QCD models, including the extended Linear sigma model (eLSM)~\cite{Eshraim:2014eka,Parganlija:2012fy} and the PNJL model~\cite{Xia:2013caa}, and are also confronted with the Particle Data Group (PDG) values~\cite{ParticleDataGroup:2022pth}.   The PDG column shows the experimental meson masses, with ground states listed first and their radial excitations added where available. Radial excitations are simply heavier partners of the ground states, carrying the same quantum numbers but reflecting a higher internal quark--antiquark mode.   The results exhibit strong consistency with experimental data and with predictions from other QCD-based effective models. In particular, the agreement between the SU($4$) PQM model and the eLSM is notably fine, especially in both the scalar and pseudoscalar meson sectors.   The extended SU($4$) PQM model not only reproduces the ground-state spectrum but also matches well with several known excitations, such as $\pi (1300), \eta(1405/1475)$ and $f_0(1370)$. This agreement suggests that the model is able to capture important aspects of chiral symmetry restoration and confinement dynamics across the light, strange and charm sectors.

Tab.~\ref{tab:meson masses} aims to estimate meson state masses with high precision, not only for vacuum phenomenology but also for applications in thermal and dense QCD media. The available fits in the SU($3$) and SU($4$) PQM models are obtained from the expressions for the squared tree-level masses listed in Tabs.~\ref{PSTab1} and \ref{PSTab2}, corresponding to the pseudoscalar and scalar sectors, respectively. The estimated fits of nonet meson states in the SU($3$) framework assume a vanishing charm condensate $\sigma_c$. The SU($3$) PQM model accurately predicts the $\pi$ and $K$ meson masses, showing excellent agreement with PDG values (within $1$--$2\%$), thereby confirming the effectiveness of the chiral Lagrangian in describing the Nambu--Goldstone bosons. However, it significantly underestimates the $\eta$ and $\eta^\prime$ masses, revealing limitations in capturing the $U(1)_A$ anomaly within the SU($3$) framework. Moreover, the states $\eta_\ell,\; \eta_s,\; \sigma_\ell$ and $\sigma_s$ are model basis states (light- and strange-quark isosinglets). The predicted masses of $\eta_\ell(1342)$ and $\eta_s(1469)$ correspond closely to the experimental $\eta(1408)$ and $\eta(1475)$ states, respectively. While the $\sigma_\ell (1458.01)$ and $\sigma_s (1671.23)$ states are identified with the experimental $\sigma (1370)$ and $\sigma(1506)$ states.

\begin{figure}[htb]
\centering{
\includegraphics[width= 16.5 cm, height=5.5 cm, angle=0  ]{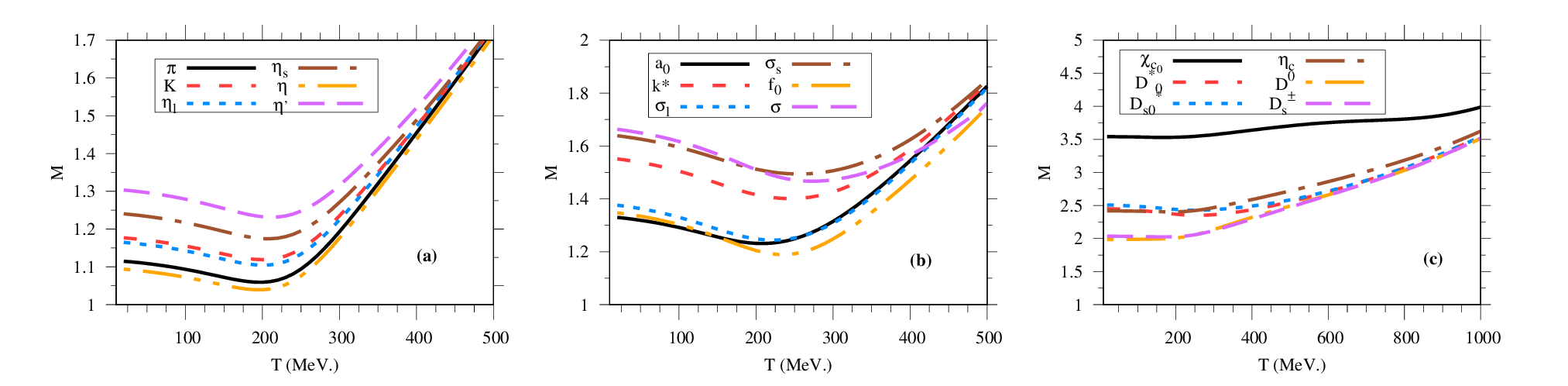}
\caption{(Color online) In-medium meson masses for the nonet pseudoscalar (left panel), scalar (middle panel), and open/hidden charmed meson sectors (right panel).
}
\label{Fig-Masses}} 
\end{figure}

In the SU($4$) PQM model, the focus shifts to the neutral multiplet masses of nonstrange--charmed meson states. This approach employs a low-energy chiral model in which the input parameters directly determine the calculated meson masses. The extended flavor symmetry enables exploration of charm-sector phenomenology while maintaining consistency with the light-quark sector. The choice of fitting parameters plays an essential role in achieving agreement with PDG values and results from other low-energy effective models. In the present work, the parameters are fixed at $\sigma = 800$~MeV, as specified in Ref.~\cite{Schaefer:2008hk}. In the presence of chiral symmetry breaking, isospin asymmetry among light quarks, the anomaly term, and the Polyakov-loop corrections, Fig.~\ref{Fig-Masses} presents the masses of nonet meson sectors: pseudoscalar (negative parity) and scalar (positive parity) meson states (left and middle panels), along with open and hidden charmed mesons in the thermal QCD medium (right panel).

\begin{table}[htb]
\centering
\begin{tabular}{l|c|c|c|c|c}
\toprule
\textbf{Meson} & \begin{tabular}[c]{@{}c@{}} \textbf{PDG}  \cite{ParticleDataGroup:2022pth} \\  (ground; excitation) \end{tabular} & \textbf{SU($3$) PQM} & \textbf{eLSM} \cite{Eshraim:2014eka, Parganlija:2012fy} & \textbf{PNJL} \cite{Xia:2013caa} &  \textbf{SU($4$) PQM}   \\
\midrule
\multicolumn{6}{l}{\textbf{Pseudoscalar Sector ($0^{-+}$)}} \\
\midrule
$\pi$ & \begin{tabular}[c]{@{}c@{}} $\pi^0(134.98 \pm 0.06)$ ; \\ $\pi(1300 \pm 100)$  \end{tabular} & $136.8$ & $141.0\pm5.8$ & $126$ & $1291$  \\
$K$ & \begin{tabular}[c]{@{}c@{}}  $K^0(497.61 \pm 0.01)$; \\   $K(1460)$ \end{tabular} & $496.3$ & $485.6\pm3.0$ & $490$ & $1228$   \\
$\eta_\ell$ &--&--& --& --& $1342$
\\
$\eta_s$ &--&--& --& --& $1469$ 
\\
$\eta$ & \begin{tabular}[c]{@{}c@{}}   $\eta(547.86 \pm 0.02)$;  \\ $\eta(1294 \pm 4)$  \end{tabular}   & $553$ & $509.4\pm3.0$ & $505$ & $1276$   \\
$\eta^\prime$ & \begin{tabular}[c]{@{}c@{}} $\eta^\prime(957.78 \pm 0.06)$; \\ $\eta^\prime(1475 \pm 4) $\end{tabular}     & $635$  & $962.5\pm 5.6$ & $949$ & $1527$  \\
$\eta_c$ & $\eta_c(2983.9 \pm 0.4)$  & -- & $2673\pm 118 $ & -- & $2682$   \\
$D$ & $D^0(1864.84 \pm 0.05)$ & -- & $1981\pm73$ & -- & $1985.73$   \\
$D_s$ & $D_s^\pm(1968.35 \pm 0.07)$  & -- & $2004\pm73$ & -- & $2035.02$  \\

\midrule
\multicolumn{6}{l}{\textbf{Scalar Sector ($0^{++}$)}} \\
\midrule
$a_0 $ &\begin{tabular}[c]{@{}c@{}}  $a_0(980 \pm 20)$;  \\ $a_0(1474 \pm 19)$ \end{tabular} & $849.8$ & $1636\pm1$ & $837$ & $1381.58$  \\
$\kappa$ &\begin{tabular}[c]{@{}c@{}} $k^*_0(682 \pm 29)$; \\ $k^*_0(1425 \pm 50)$ \end{tabular}  & $1331.6$ & $1450\pm 1$& $1013$ & $1581.45$  \\
$\sigma_\ell$ &--&--& --& --& $1458.01$ 
\\
$\sigma_s$ &--&--& --& --& $1671.23$ 
\\
$\sigma$ &\begin{tabular}[c]{@{}c@{}}   $400$--$1200$; \\ $\sigma(1704 \pm 12)$ \end{tabular} & $824.0$ & $975.1\pm 6.4$ & $700$ & $1645.96$  \\
$f_0$ &  $1200$--$1500$  & $1339.0$ & $1186 \pm 6$ & $1169$ & $1241.92$   \\
$\chi_{c_0}(1P)$ &$\chi_{c_0}(3414.71 \pm 0.3)$ & -- & $3144\pm128$ & -- & $3541.21$   \\
$D_0^*$ & $D_0^*(2343.00 \pm 10.0)$ & -- & $2414\pm77$ & -- & $2451.11$  \\
$D_{s0}^*$ &  $D_{s_0}^* (2317.8 \pm 0.5)$ & -- & $2467\pm76$ & -- &$ 2507.47  $\\
\bottomrule

\end{tabular}
 \caption{ A comparison between (pseudo)scalar meson states within PQM model. \label{tab:meson masses}} 
\end{table}

Figure~\ref{Fig-Masses} illustrates the thermal evolution of the meson mass spectrum for the nonet pseudoscalar mesons ($\pi$, $K$, $\eta_\ell$, $\eta_s$, $\eta$, $\eta^\prime$) (left panel), nonet scalar mesons ($a_0$, $\kappa$, $\sigma_\ell$, $\sigma_s$, $f_0$, $\sigma$) (middle panel), and open/hidden charmed mesons ($D^0$, $D^\pm$, $D_0^*$, $D_{s0}^*$, $\eta_c$, $\chi_{c0}$) (right panel), all given at vanishing chemical potential $\mu_f$. The temperature dependence of the meson mass spectrum can be categorized into three regions: the bosonic (mesonic) region, the phase transition region, and the fermionic (partonic) region. In the bosonic region, meson masses decrease gradually with increasing temperature, reflecting the dominance of thermal suppression of mesonic degrees of freedom. The phase transition region is characterized by a smooth but rapid decline in mass values, corresponding to the chiral crossover. At high temperatures, in the fermionic region, quark--antiquark contributions dominate, leading to a moderate increase in the meson masses due to the restoration of chiral symmetry.

At low temperatures ($T < T_{\chi}$), bosonic contributions dominate. The tree-level meson masses presented in Tables~\ref{PSTab1} and~\ref{PSTab2}  effectively restore the mass gap characteristic of the chiral phase structure in meson states. As previously discussed, the system remains in the confined phase, characterized by large chiral condensates $\sigma_f$ and small values of the Polyakov-loop order parameters. This regime allows for the estimation of vacuum mass values for each meson sector at vanishing temperature ($T = 0~\mathrm{MeV}$). In this region, meson masses are expressed as functions of pure mesonic condensates $\sigma_f$. The thermal behavior of meson masses remains nearly constant, relative to their corresponding vacuum values, until the temperature approaches the pseudo-critical point $T \sim T_{\chi}$. It is evident that in this low-temperature regime, the masses of all meson states are effectively independent of temperature.
 
The $U(1)_A$ symmetry breaking appears in the tree-level meson masses through the anomaly term $\mathcal{C}$, which is particularly significant for nonet meson states (left and middle panels) due to strange and charm quark contributions. In the SU($4$) PQM model, the phase structure of the charm chiral condensate $\sigma_c$ plays an essential role in determining the chiral behavior of meson masses. When $\sigma_c$ vanishes, the SU($3$) structure is effectively restored. In this limit, the $U(1)_A$ anomaly effect disappears in the charm--strange meson sectors, and the term $\mathcal{L}_{\mathrm{emass}}$ becomes dominant. This term is directly related to the explicit quark mass contribution. For simplicity, we assume $\varepsilon_u = \varepsilon_d = \varepsilon_s = 0$ and $\varepsilon_c \sim m_c$.

In the phase transition region, beyond the chiral temperature $T \sim T_{\chi}$, meson masses gradually decrease with increasing temperature. This region exhibits a smooth decline, characteristic of a slow crossover transition. Beyond the chiral temperature ($T \sim T_{\chi}$), meson masses gradually decrease with rising temperature. This behavior reflects the nature of a smooth crossover transition. Notably, the location of the chiral critical endpoint (CEP) is not universal across all meson states; rather, it depends on the specific PQM order parameters and the input conditions of the QCD medium.

At high temperatures ($T > T_{\chi}$), fermionic (partonic) thermal contributions significantly influence the thermal evolution of meson states. As the temperature increases, these contributions become dominant and lead to the degeneration of meson masses. In contrast, their impact remains negligible at low temperatures. The observed degeneracy in meson masses arises from thermal fluctuations acting on the chiral condensates $\sigma_f$, particularly the strange $\sigma_s$ and charm $\sigma_c$ condensates, and reflects the process of chiral symmetry restoration~\cite{Schaefer:2008hk}. These thermal fluctuations melt the light quark condensates ($\sigma_u$, $\sigma_d$) more rapidly than the strange and charm condensates. Consequently, nonet pseudoscalar mesons (left panel) enter the deconfined (partonic) phase earlier than nonet scalar and charmed mesons. At very high temperatures ($T \sim 1~\mathrm{GeV}$), the thermal evolution of the charm condensate completes the transition for strange--charm meson states. In this regime, charmed meson masses remain relatively stable until the temperature becomes extremely high. Accordingly, charmed meson states are expected to undergo phase transition and mass degeneracy as the density of the system increases. This transition enhanced thermal energy to overcome the gap associated with the  Fermi surface of the meson states  \cite{Tawfik:2014gga,Tawfik:2019rdd}, This latter differs from one meson state to another.

\section{Conclusions \label{conclusion}}
In conclusion, we extended the SU($4$) Polyakov Quark Meson (PQM) model by incorporating isospin asymmetry to explore the chiral phase structure of QCD matter in a thermal medium. This study examined how the system undergoes phase transitions in the presence of isospin asymmetry between light quark condensates and the charm condensate. The results reveal key features of chiral symmetry restoration, deconfinement transitions, quark and charm fluctuations and correlations, thermodynamic observables, and the thermal evolution of meson masses within the SU($4$) PQM framework.

The introduction of isospin asymmetry leads to several modifications in the chiral phase structure and meson dispersion relations. At finite isospin chemical potential ($\mu_I$), the $u$- and $d$-quark condensates ($\sigma_u$, $\sigma_d$) split due to isospin symmetry breaking. In the SU(4) framework, all diagonal sigma-fields ($\sigma_0$, $\sigma_3$, $\sigma_8$, $\sigma_{15}$) and symmetry-breaking parameters ($h_0$, $h_3$, $h_8$, $h_{15}$) become nonzero, explicitly breaking SU(2) isospin symmetry through the presence of $\sigma_3$ and $h_3$. As a result, the pseudo-critical temperatures for $\sigma_u$ and $\sigma_d$ decrease with increasing $\mu_I$.

The inclusion of the charm chiral condensate ($\sigma_c$) significantly enhances the thermal resolution of the PQM  model, particularly at high temperatures. While light and strange quark condensates melt rapidly near the pseudo-critical temperature, the charm condensate remains largely stable up to much higher temperatures. This enables better characterization of the crossover and high-temperature phases. The incorporating of the isospin asymmetry allows for tracing the $(T/t_chi, \mu_I/m_\pi)$ plane, where the pseudo-critical temperature $T_\chi$ decreases with increasing normalized isospin chemical potential. The model agrees well with lattice QCD simulations for $\mu_I/m_\pi < 1.0$, and the inclusion of $\sigma_c$ enables exploration of high-temperature and high-density regimes.

Thermodynamic observables-such as pressure, trace anomaly, and entropy density--calculated from the SU($4$) PQM model show excellent agreement with lattice QCD results. All quantities exhibit a smooth crossover and remain below the Stefan--Boltzmann (SB) limit at high temperatures. These results are critical for modeling the EoS for dense QCD matter relevant to compact astrophysical objects. While $\sigma_c$ may play a minor role in early-stage deconfinement, it could influence the EoS in the extreme interior of hybrid stars. The analysis of diagonal susceptibilities $\chi_f^2$ for $f = (\ell, s, c)$ provides insight into the thermal behavior of the corresponding chiral condensates. Light and strange quark susceptibilities rise sharply near $T_\chi$, while the charm susceptibility increases gradually, indicating delayed thermal response in the charm sector.

Meson masses were obtained from the second derivative of the SU($4$) PQM grand potential with respect to the mesonic fields. This includes tree-level, thermal, and vacuum contributions, allowing for a consistent description of the meson mass spectrum. Light pseudoscalar mesons such as $\pi$, $K$, and $\eta$ show rapid mass suppression with increasing temperature, while scalar mesons such as $a_0$, $\kappa$, and $f_0$ exhibit greater thermal stability. The inclusion of $\sigma_c$ enables analysis of the thermal behavior of both open-charm mesons (e.g., $D$, $D_s$) and hidden charmonia (e.g., $\eta_c$, $\chi_{c0}$), which remain thermally stable over a wide temperature range.

In dense astrophysical systems such as neutron stars and hybrid stars, where both baryon density and temperature are extreme, charm mesons may contribute to the EoS. The slow melting of $\sigma_c$ and the thermal robustness of hidden charmed mesons suggest potential relevance for transport and stability properties in such studies. Overall, the SU($4$) PQM model provides a well framework for studying the QCD phase diagram, including heavy-flavor dynamics. It offers predictive power for both heavy-ion experiments at FAIR and NICA and for modeling extreme conditions in compact stars.

\section*{Acknowledgements}
The author gratefully acknowledges the anonymous reviewer for their invaluable feedback and constructive suggestions, which have significantly improved the quality of this work.

 \appendix 
 \section{SU(4) Generator Operators and Meson Structure \label{AppA}}

The generator operator $\hat{T}_a = \hat{\lambda}_a / 2$ in U(4) is obtained from the generalized Gell-Mann matrices $\hat{\lambda}_a$~\cite{Weinberg:1972kfs}, where the indices run as $a = 0, \ldots, 15$ \cite{Lenaghan:2000ey}. The  matrices are defined as a follow:-
\bea
 \label{the Gell-Mann matrices}
\hat{\lambda}_{1}= \left[
\begin{array}{cccc}
0 & 1 & 0 & 0\\
1 & 0 & 0 & 0\\
0 & 0 & 0 & 0\\
0 & 0 & 0 & 0
\end{array}
\right]
,
\hat{\lambda}_{2}= \left[
\begin{array}{cccc}
0 & -i & 0 & 0 \\
i & 0 & 0 & 0\\
0 & 0 & 0 & 0\\
0 & 0 & 0 & 0
\end{array}
\right]
,
\hat{\lambda}_{3}= \left[
\begin{array}{cccc}
1 & 0 & 0 & 0\\
0 & -1 & 0 & 0\\
0 & 0 & 0 & 0\\
0 & 0 & 0 & 0
\end{array}
\right]
,
\hat{\lambda}_{4}= \left[
\begin{array}{cccc}
0 & 0 & 1 & 0\\
0 & 0 & 0 & 0\\
1 & 0 & 0 & 0\\
0 & 0 & 0 & 0
\end{array}
\right]
,\nonumber \\
\hat{\lambda}_{5}= \left[
\begin{array}{cccc}
0 & 0 & -i & 0\\
0 & 0 & 0 & 0\\
i & 0 & 0 & 0\\
0 & 0 & 0 & 0
\end{array}
\right]
, 
\hat{\lambda}_{6}= \left[
\begin{array}{cccc}
0 & 0 & 0 & 0 \\
0 & 0 & 1 & 0\\
0 & 1 & 0 & 0\\
0 & 0 & 0 & 0
\end{array}
\right]
,
\hat{\lambda}_{7}= \left[
\begin{array}{cccc}
0 & 0 & 0 & 0\\
0 & 0 & -i & 0\\
0 & i & 0 & 0\\
0 & 0 & 0 & 0
\end{array}
\right]
,
\hat{\lambda}_{8}= \frac{1}{\sqrt{3}} \left[
\begin{array}{cccc}
1 & 0 & 0 & 0\\
0 & 1 & 0 & 0\\
0 & 0 & -2 & 0\\
0 & 0 & 0 & 0
\end{array}
\right]
,\nonumber \\
\hat{\lambda}_{9}= \left[
\begin{array}{cccc}
0 & 0 & 0 & 1\\
0 & 0 & 0 & 0\\
0 & 0 & 0 & 0\\
1 & 0 & 0 & 0
\end{array}
\right]
,
\hat{\lambda}_{10}= \left[
\begin{array}{cccc}
0 & 0 & 0 & -i\\
0 & 0 & 0 & 0\\
0 & 0 & 0 & 0\\
i & 0 & 0 & 0
\end{array}
\right]
, 
\hat{\lambda}_{11}= \left[
\begin{array}{cccc}
0 & 0 & 0 & 0\\
0 & 0 & 0 & 1\\
0 & 0 & 0 & 0\\
0 & 1 & 0 & 0
\end{array}
\right]
,
\hat{\lambda}_{12}= \left[
\begin{array}{cccc}
0 & 0 & 0 & 0\\
0 & 0 & 0 & -i\\
0 & 0 & 0 & 0\\
0 & i & 0 & 0
\end{array}
\right]
,\nonumber \\
\hat{\lambda}_{13}= \left[
\begin{array}{cccc}
0 & 0 & 0 & 0\\
0 & 0 & 0 & 0\\
0 & 0 & 0 & 1\\
0 & 0 & 1 & 0
\end{array}
\right]
,
\hat{\lambda}_{14}= \left[
\begin{array}{cccc}
0 & 0 & 0 & 0\\
0 & 0 & 0 & 0\\
0 & 0 & 0 & -i\\
0 & 0 & i & 0
\end{array}
\right]
,
\hat{\lambda}_{15}= \frac{1}{\sqrt{6}}\left[
\begin{array}{cccc}
1 & 0 & 0 & 0\\
0 & 1 & 0 & 0\\
0 & 0 & 1 & 0\\
0 & 0 & 0 & -3
\end{array}
\right].
\eea

These generators satisfy the U($4$) algebra with the commutation and anticommutation relations:
\bea
[\hat{T}_a, \hat{T}_b] = i f_{abc} \hat{T}_c, \qquad
\{\hat{T}_a, \hat{T}_b\} = i d_{abc} \hat{T}_c,
\eea
where $f_{abc}$ and $d_{abc}$ are the antisymmetric and symmetric structure constants of SU($4$), respectively. The symmetric constants are explicitly defined by:
\bea
d_{abc} = \frac{1}{4} \mathrm{Tr} \left( \{ \hat{\lambda}_a, \hat{\lambda}_b \} \hat{\lambda}_c \right), \qquad
d_{ab0} = \sqrt{\frac{2}{3}} \delta_{ab}.
\eea
In the framework of PCAC, the decay constant $f_a$ is related to the symmetric constants and sigma fields as:
\bea
f_a = d_{aab} \bar{\sigma}_a.
\eea
For charged and neutral pion meson ($f_{\pi^{\pm}} =f_1,\;f_{\pi^0} =f_3$) and kaon meson ($f_{K^{\pm}} =f_4,\;f_{K^0} =f_6$)  are given as , the decay constants take the form:
\bea
f_{\pi^0} &=& f_{\pi^\pm} = \sqrt{\frac{2}{3}} \bar{\sigma}_0 + \frac{1}{\sqrt{3}} \bar{\sigma}_8, \\
f_{K^\pm} &=& \sqrt{\frac{2}{3}} \bar{\sigma}_0 + \frac{1}{2} \bar{\sigma}_3 - \frac{1}{2\sqrt{3}} \bar{\sigma}_8, \\
f_{K^0} &=& \sqrt{\frac{2}{3}} \bar{\sigma}_0 - \frac{1}{2} \bar{\sigma}_3 - \frac{1}{2\sqrt{3}} \bar{\sigma}_8.
\eea
Hence, the isospin field $\bar{\sigma}_3$ can be expressed as:
\bea
\bar{\sigma}_3 = f_{K^\pm} - f_{K^0}.
\eea
The scalar meson generator matrix $T_a \sigma_a$ takes the following $4 \times 4$ matrix form:
\bea
T_a \sigma_a = \frac{1}{\sqrt{2}} \begin{pmatrix}
\frac{\sigma_0}{2}+\frac{\sigma_3}{\sqrt{2}}+\frac{\sigma_8}{\sqrt{6}}+\frac{\sigma_{15}}{2\sqrt{3}} &\frac{\sigma_1-i\sigma_2}{\sqrt{2}}&\frac{\sigma_4-i\sigma_5}{\sqrt{2}}& \frac{\sigma_9-i\sigma_{10}}{\sqrt{2}}\\
\frac{\sigma_1+i\sigma_2}{\sqrt{2}} &\frac{\sigma_0}{2}-\frac{\sigma_3}{\sqrt{2}}+\frac{\sigma_8}{\sqrt{6}}+\frac{\sigma_{15}}{2\sqrt{3}}& \frac{\sigma_6-i\sigma_{7}}{\sqrt{2}} & \frac{\sigma_{11}-i\sigma_{12}}{\sqrt{2}}\\
\frac{\sigma_4+i\sigma_5}{\sqrt{2}}&  \frac{\sigma_6+i\sigma_{7}}{\sqrt{2}} &\frac{\sigma_0}{2} -\frac{2}{\sqrt{6}} \sigma_8 +\frac{\sigma_{15}}{2\sqrt{3}} & \frac{\sigma_{13}-i\sigma_{14}}{\sqrt{2}}\\ 
\frac{\sigma_9+i\sigma_{10}}{\sqrt{2}} &\frac{\sigma_{11}+i\sigma_{12}}{\sqrt{2}} & \frac{\sigma_{13}+i\sigma_{14}}{\sqrt{2}} & \frac{\sigma_0}{2} -\frac{\sqrt{3}}{2} \sigma_{15}
\end{pmatrix}.
\eea
Similarly, the pseudoscalar meson generator matrix $T_a \pi_a$ becomes:
\bea
T_a \pi_a = \frac{1}{\sqrt{2}} \begin{pmatrix}
\frac{\pi_0}{2}+\frac{\pi_3}{\sqrt{2}}+\frac{\pi_8}{\sqrt{6}}+\frac{\pi_{15}}{2\sqrt{3}} &\frac{\pi_1-i\pi_2}{\sqrt{2}}&\frac{\pi_4-i\pi_5}{\sqrt{2}}& \frac{\pi_9-i\pi_{10}}{\sqrt{2}}\\
\frac{\pi_1+i\pi_2}{\sqrt{2}} &\frac{\pi_0}{2}-\frac{\pi_3}{\sqrt{2}}+\frac{\pi_8}{\sqrt{6}}+\frac{\pi_{15}}{2\sqrt{3}}& \frac{\pi_6-i\pi_{7}}{\sqrt{2}} & \frac{\pi_{11}-i\pi_{12}}{\sqrt{2}}\\
\frac{\pi_4+i\pi_5}{\sqrt{2}}&  \frac{\pi_6+i\pi_{7}}{\sqrt{2}} &\frac{\pi_0}{2} -\frac{2}{\sqrt{6}} \pi_8 +\frac{\pi_{15}}{2\sqrt{3}} & \frac{\pi_{13}-i\pi_{14}}{\sqrt{2}}\\ 
\frac{\pi_9+i\pi_{10}}{\sqrt{2}} &\frac{\pi_{11}+i\pi_{12}}{\sqrt{2}} & \frac{\pi_{13}+i\pi_{14}}{\sqrt{2}} & \frac{\pi_0}{2} -\frac{\sqrt{3}}{2} \pi_{15}
\end{pmatrix}.
\eea


 \bibliographystyle{aip}

\bibliography{PQMREF.bib}  
 
\end{document}